%% file: wiki_umich_www25.tex
\documentclass{article}

\usepackage[a4paper, margin=1in]{geometry} 
\usepackage{graphicx}   
\usepackage{amsmath, amssymb} 
\usepackage{hyperref}   
\usepackage{natbib}     

\usepackage[utf8]{inputenc} 
\usepackage[T1]{fontenc}    
\usepackage{hyperref}       
\usepackage{url}            
\usepackage{booktabs}       
\usepackage{amsfonts}       
\usepackage{nicefrac}       
\usepackage{microtype}      
\usepackage{lipsum}
\usepackage{fancyhdr}       
\usepackage{graphicx}       
\graphicspath{{media/}}     

\usepackage{subfig}
\usepackage{multirow}   
\usepackage{arydshln} 
\usepackage{bbm} 
\usepackage{enumitem} 
\usepackage{amsmath}
\usepackage{etoolbox}
\usepackage{fancyhdr}
\usepackage{amssymb}

\newcommand{\ignore}[1]{}

\newcommand{\dataset}{\textsl{SustainPedia}}
\newcommand{\Hquad}{\hspace{0.5em}}

\title{A Test of Time: Predicting the Sustainable Success of Online Collaboration in Wikipedia}
\author{
    Abraham Israeli$^{1,*}$, David Jurgens$^1$, and Daniel Romero$^1$ \\  
    \small $^1$University of Michigan, School of Information (UMSI), Ann Arbor, MI, USA\\  
    \small $^*$Email: isabrah@umich.edu
}
\date{}

\begin{document}

\maketitle

\begin{abstract}
The Internet has significantly expanded the potential for global collaboration, allowing millions of users to contribute to collective projects like Wikipedia. While some of these collaborative efforts see early success, their long-term success is key to creating lasting impact. Despite its importance, this dimension of success remains largely unexplored. Prior work has assessed the success of online collaborations, however, most approaches are time-agnostic, evaluating success without considering its longevity.
Research on the factors that ensure the long-term preservation of high-quality standards in online collaboration is scarce. In this study, we address this gap. We propose a novel metric, `Sustainable Success,' which measures the ability of collaborative efforts to maintain their quality over time. 
Using Wikipedia as a case study, we introduce the \dataset~dataset, which compiles data from over 40K Wikipedia articles, including each article's sustainable success label and more than 300 explanatory features such as edit history, user experience, and team composition. 
Using this dataset, we develop machine learning models to predict the sustainable success of Wikipedia articles.
Our analysis reveals important insights. For example,  we find that the longer an article takes to be recognized as high-quality, the more likely it is to maintain that status over time (i.e., be sustainable). Additionally, user experience emerged as the most critical predictor of sustainability.
Our analysis provides insights into broader collective actions beyond Wikipedia, such as online activism and crowdsourced open-source software, where the same social dynamics that drive success on Wikipedia might play a role.
\end{abstract}

\section{Introduction}
\label{sec:intro}

Collaboration plays a crucial role in societal progress, shaping the way individuals and groups share knowledge, solve problems, and innovate \cite{sachs2004evolution, salminen2012collective, ostrom1990governing, tschannen2001collaboration}. In the digital era, platforms like Wikipedia and GitHub have revolutionized collaboration by allowing millions of users to collectively create content \cite{ebersbach2008wiki, godwin2003blogs, lima2014coding, dabbish2012social, piccardi2021value}. However, while collaboration is rooted in human behavior, and online platforms empower millions to do so easily, the reality remains that not all collaborative efforts lead to \emph{success}.

Humans have long sought to assess group success across various dimensions and contexts \cite{hackman2002leading,woolley2010evidence, pentland2012new,cunha2019all,kim2006community,kraut2012building}. In particular, evaluating and predicting the success of collaborative groups is important for optimizing work efficiency and gaining deeper insights into human dynamics.

As a result, there has been growing interest in understanding the factors that determine the success or failure of crowd collaboration. Research in this area is primarily motivated by two key objectives: (i) Better predicting the success of online collaborations \cite{moas2023automatic}; and (ii) Uncovering the underlying mechanisms and factors that foster success \cite{platt2018network, israeli2024flying, luther2010works}. Advancing this knowledge is essential not only for improving existing collaborative efforts but also for the management, moderation, and design of future collaborative systems that maximize engagement, efficiency, and impact.

However, most proposed approaches focus on success at a single point in time, overlooking the importance of \textit{sustained} success over the long term, which necessitates continuous effort and ongoing maintenance by the collaborative team. While the importance of maintenance, repair, and long-term care has recently been recognized in innovation studies \cite{vinsel2020innovation}, it has yet to be fully explored in the context of collaborative efforts. 

Sustained success in collaborative environments is a challenge. Early signals of success do not always guarantee long-term viability. Prior work explored factors that influence the persistence of collaborative efforts, such as governance structures \cite{forte2009decentralization}, contributor retention \cite{halfaker2013rise, cascaes2017onboarding}, and social dynamics within online groups \cite{kittur2007power, kraut2014role}. However, none of these studies have explored the capacity of collaborative efforts to maintain a high-quality status over time---i.e., avoid an initially successful collaboration falling apart. To address this gap, we introduce a novel metric called \emph{'Sustainable Success,'} which assesses this aspect of success directly -- the ability of collaborative teams to maintain high-quality standards over time.

Using the new sustainable success measure, this study asks the following two research questions:\\
(RQ1) Can the sustainable success of an online collaboration be accurately predicted?; and\\
(RQ2) What factors contribute to the sustainable success of an online collaboration?

To answer these questions, we measure collaborations on over a decade of behavior in Wikipedia.
Through collaborative efforts, Wikipedia's content stands as one of the largest free knowledge repositories in the world \cite{similarweb2024websites, ren2023did}. Wikipedia has a rich collaborative environment, with various means of user interaction, including voting, discussions, and communication. Crucially, Wikipedia editors also directly evaluate its content in a structured way \cite{hu2007measuring, stvilia2005information, warncke2013tell, zhang2017crowd}, providing a reliable basis for measuring sustainable success. The public accessibility of Wikipedia’s data allows us to comprehensively explore the factors that contribute to the sustainability of collaborative efforts, thereby answering our RQs in a thorough and reliable way.

In our study, we use Wikipedia's two most trustworthy content assessment levels: Featured Article ($FA$) and Good Article ($GA$). The two tags signify high-quality articles that meet specific editorial standards \cite{wikipedia2024FA, wikipedia2024GA}. $FA$ represents the pinnacle of quality, having undergone rigorous peer review and meeting the strictest criteria, such as comprehensive coverage, neutrality, and well-researched content.

$GA$s, while not as rigorous as $FA$s, also meet substantial quality standards, including being well-written, factually accurate, and verifiable.
As of 02/2025, just 0.1\% and 0.49\% of English Wikipedia articles are tagged as $FA$ and $GA$, respectively. $FA$s are marked with a bronze star symbol while $GA$s with a green plus. Both tags highlight Wikipedia's best work, reflecting reliable, in-depth content.

\begin{figure}
    \centering
    \includegraphics[scale=0.45]{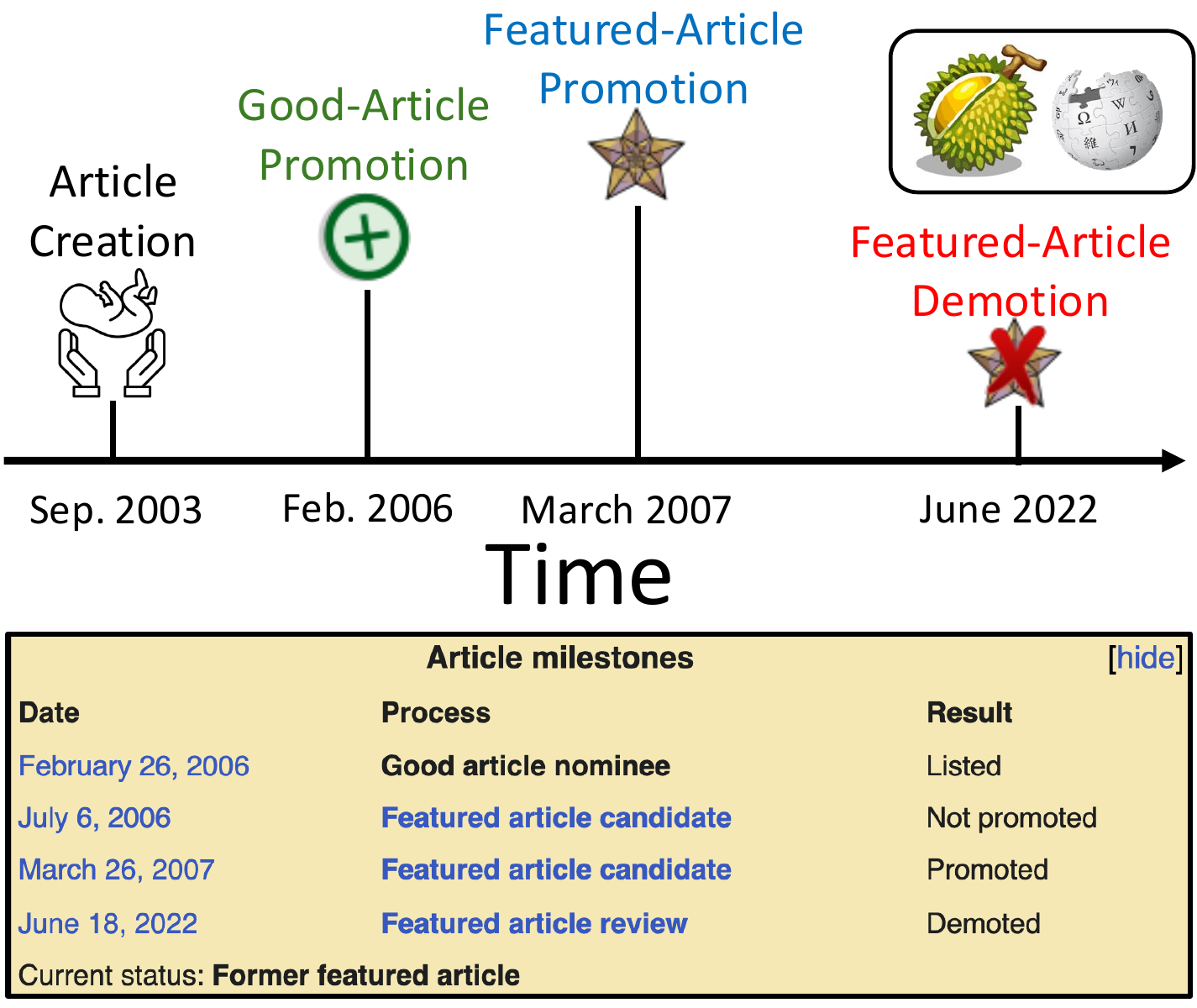}
    \caption{Article life-cycle example. (Top) Major events of the Wikipedia article about the edible fruit Durian. (Bottom) A screenshot from the talk page of the Durian article, including both its promotions and demotion. This article is tagged as unsustainable in \dataset~due to its demotion in 2022.}
    \label{fig:article_life_cycle}
\end{figure}

The community's judgment of quality is not permanent, however.
The $FA$ and $GA$ statuses can both be removed if the article no longer meets the community's quality standards, effectively demoting the article. Figure \ref{fig:article_life_cycle} illustrates this promotion and demotion process. We leverage this ongoing quality assessment to model the sustainable success of Wikipedia articles.

To understand the process behind sustained success, our models use different categories of factors related to creating and maintaining the article. Specifically, we build the following six feature sets: network structure of collaborators, topics that the article is associated with, team composition, discussion patterns, edit history, and the experience of the article's collaborators.

To summarise, our main three contributions are:
\begin{enumerate}[leftmargin=0.2cm]
    \item We introduce a novel definition and measurement of continued collaborative success. We define it as the ability to keep high standards over time, which we name as \emph{sustainable success}.
    \item We present a good-performing machine learning model to predict the sustainable success of Wikipedia articles and identify the factors that are most associated with sustainability, such as the experience of team members.
    \item We introduce a new dataset named \dataset~containing all the information we used for analysis and modeling. We make to \dataset~public to be further used by the research community. 
\end{enumerate}

\section{Online Collaboration and Success}
\label{sec:online_collaboration_and_success}
Numerous researchers investigate collaborative behaviors on social platforms like Reddit and X (formerly Twitter) \cite{honey2009beyond, weninger2013exploration, stoddard2015popularity, panek2018effects, mensah2020characterizing}. Additionally, some focus on identifying these collaborative patterns within specialized crowd-sourced platforms, including Q\&A websites, open-source initiatives, and Wikipedia \cite{wang2013empirical, ransbotham2011membership, dabbish2012social, keegan2012editors, thung2013network, zhang2017crowd}. 

A significant area of focus in this field is quantifying and predicting the success of online activities \cite{tan2018tracing, cunha2019all, israeli2024flying}, particularly in relation to crowd collaboration \cite{das2022quality,das2024language}. Many of such efforts focus on understanding current algorithms used for quality assessment in Wikipedia \cite{ren2023did}. These success metrics, among other things, help identify the elements that contribute to successful online collaboration such as edit behavior and group dynamics.


In the past decade, numerous studies have demonstrated how factors such as edit behavior \cite{liu2011does, ruprechter2020relating}, network structure \cite{raman2020classifying,brandes2009network,lerner2019network,de2015measuring}, regulation \cite{ruprechter2023protection}, and textual and visual content \cite{bassani2019automatically,shen2017hybrid,shen2019joint,guda2020nwqm} are associated with content assessment in Wikipedia. Some of those suggest advanced deep learning models that integrate multiple data sources using a multimodal architecture.

The most closely related work to ours is by \citet{das2022quality}. The authors proposed an automated approach for the detection of the early signals influencing the assessment level
change of Wikipedia articles. Their approach models an article as a time series of consecutive revisions, with each revision represented by a set of features to predict future quality levels. Our research draws inspiration from their work and offers extensions in the following significant ways: (i) We emphasize both the predictive power and interpretability of our proposed model; (ii) Unlike \citet{das2022quality}, who examined a small subset of English Wikipedia articles, we include \emph{all} relevant articles in our study; and (iii) \citet{das2022quality} use all existing quality assessment levels in Wikipedia. However, we deem some of those levels as less reliable. Therefore, we limit our focus to the $FA$ and $GA$ quality assessment levels, which are established through an official review process and a broad consensus by the Wikipedia community. We discuss this further in Section \ref{subsec:wikipedia_quality_assessment_system}

\section{Problem Definition}
\label{sec:problem_def}
We investigate the sustainability of successful online collaborations by focusing on the sustainable success of English Wikipedia articles. English Wikipedia maintains a well-structured and documented quality assessment system in which each article is assessed with its quality level(s).

\paragraph{Notation} We use the following notation in the rest of the paper. $A$ is the set of Wikipedia articles we process. For each article $a \in A$, we mark to the following three timestamps: $t_{birth}^a$, $t_{prom}^a$, $t_{dem}^a$. The first refers to the time when the article was created (i.e., its first revision), and the other two refer to the time when the article was promoted and demoted, respectively. Some articles lack the $t_{dem}^a$ timestamp as they were never demoted.

\subsection{Wikipedia's Quality Assessment System}
\label{subsec:wikipedia_quality_assessment_system}

Articles are assigned quality scores in Wikipedia in two ways. First, Wikipedia maintains two quality assessment classes where an article can be tagged as ``Featured Article'' ($FA$) or ``Good Article'' ($GA$). $FA$ and $GA$ determinations are made using a Wikipedia broad consensus. $FA$s are ``the best articles Wikipedia has to offer'' \cite{wikipedia2024FA} while $GA$s meet ``a core set of editorial standards'' \cite{wikipedia2024GA} but are not featured article quality. 
Second, quality grades (e.g., B, C, Start) are assigned by \emph{WikiProjects}, which are groups of thematically related articles, such as the `Plants' Wikiproject. Most articles are associated with at least one WikiProject who each assesses an article's quality separately using the same grading scale. However, each WikiProject can have different quality standards, grading systems, responsiveness to changes in an article, etc. It is common for an article to have different quality grades across WikiProjects \cite{wikipediaEngine}.

In contrast, $FA$ and $GA$ quality assessments are judged Wikipedia-wide using a systematic peer-review evaluation, which excludes the editors of the evaluated article, making them more reliable than WikiProjects scores. Thus, in this study, we only use $FA$ and $GA$ quality assessment levels. Currently, there are 6,583 and 40,191 $FA$s \cite{wikipedia2024FA} and $GA$s \cite{{wikipedia2024GA}} respectively.


%


\subsection{Promotions and Demotion}
\label{subsec:promotions_demotions}
Interestingly, both $FA$ and $GA$ quality assessment levels are dynamic. It is common for an article to first be promoted to $GA$ and then later to $FA$. Additionally, promoted articles can be demoted (and later be repromoted, and so on). As illustrated in Figure \ref{fig:article_life_cycle}, the Durian article was first promoted to $GA$, then to $FA$ before being demoted in June 2022. Demoted articles are reassigned with an assessment level of Start, C, B, or A (i.e., $FA$s are demoted to a lower level than $GA$). The main reasons to demote an article are incomplete/updated content, as new information becomes available, or the article no longer meeting Wikipedia's criteria for $FA$ or $GA$.

The demotion of an article, the same as promotion, is determined through a consensus derived through discussion at a dedicated Wikipedia discussion page. The reevaluation of an article's quality is done through an `article review' process, which usually takes a few weeks \cite{wikipedia2024FAdemotion, wikipedia2024GAdemotion}. During this time, the members of the relevant community are allowed to comment on the required improvements and directly edit the article to improve it. Appendix Figure \ref{fig:durian_review_alert} illustrates how a review process is announced. A bold comment appears on the `Talk Page' of the relevant article, asking the Wikipedia community to comment about the issue and motivating Wikipedians to edit the article and improve its content. 

In this research, we use the promotion/demotion processes in Wikipedia to learn what makes successful collaborations sustainable. We define sustainability as maintaining high-quality standards over time. In Wikipedia, this means \emph{not} being demoted after an initial promotion. For the reasons mentioned above, we only rely on the $FA$ and $GA$ quality assessment levels. 

\subsection{Binary Classification}
\label{subsec:binary_classification}
We are interested in predicting if a promoted article, $a \in A$ is \emph{unsustainable}. That is, we question if $a$ will be demoted from its high-level quality assessment level or keep its promoted status. If $a$ does not keep its promoted status, we refer to it as an unsustainable article. Hence, we define the following indicator as our target feature:

\begin{equation}
    \mathbbm{1}_{unsustain}^{a} \triangleq
    	\begin{cases}
    	1 & \text{$a$ was promoted and later demoted}\\
    	0 & \text{$a$ was promoted and never demoted}
    	\end{cases}
\label{eq:class_indic}
\end{equation}

We set the positive class to unsustainable (demoted) articles since this is the minority class and the more challenging for a classifier to predict. We define two independent classification use cases for $FA$ and $GA$ and therefore define $\mathbbm{1}_{unsustain}^{a}$ per use case independently. Our definition has the following properties:
\begin{itemize}[leftmargin=0.2cm]
    \item \textit{Demotion}: We tag an article as unsustainable if it was demoted \emph{at least once} in its lifecycle. In rare cases, articles are repromoted after being demoted. However, our research objective is to find what makes an article sustainable over time, and hence, any demotion makes the article unsustainable even if it was later repromoted.
    \item \textit{From $GA$ to $FA$}: Many of the $FA$s are first promoted to a $GA$ status (see Figure \ref{fig:article_life_cycle}). These articles are being used to model both use cases -- $FA$ and $GA$. In cases where the article is later demoted, this demotion is ``counted'' for both use cases. For example, the Durian article illustrated in Figure \ref{fig:article_life_cycle} is tagged as unsuitable for both the 
    $FA$ and the $GA$ prediction models. 
    
\end{itemize}

\section{Data}
\label{sec:data}
Each Wikipedia article maintains its content page (the actual article) and its talk page. In this research, we make use of both pages. While the content page and its meta-data contain essential information on how the article's content is built and by whom, the latter contains discussions between the community members. We collect the data of all relevant articles using the Wikipedia dump files, which are monthly released by the WikiMedia Foundation \cite{wikimedia2024dumps}. 

The Wikipedia dump files allow us to collect historical information about each article. This way, we can track and collect information about each revision of both the content page and the talk page. The dump files contain detailed information per revision, including time, editor name, textual content, etc. We use the 09/23 release of the English Wikipedia.

Most talk pages contain helpful information about the article. Among those are the ``Article Milestones'' that include $FA$ or $GA$ promotions and demotions, as well as the WikiProjects associated with the article. 
We collect data for both $FA$ and $GA$ use cases in the same way to build the \dataset~dataset. In the rest of this section, we explain how we construct the \dataset~population, the target feature, and the explanatory features.

\subsection{\dataset~Population}
\label{subsec:dataset_population}

\begin{table}
\centering
\normalsize 
\caption{\dataset~dataset statistics. $\mathbbm{1}_{unsustain}$ is the classification target feature. `Prom. Time' is $t_{prom} - t_{birth}$, measured in days. Comments are taken from discussions on talk pages. The mean and standard deviation are calculated over articles. Blanks are irrelevant.}
{
    \begin{tabular}{l@{\Hquad}|c@{\Hquad}c@{\Hquad}c@{}|c@{\Hquad}c@{\Hquad}c@{}}
      \multicolumn{1}{c}{} & \multicolumn{3}{c}{\boldmath{\textbf{F}eatured \textbf{A}rticles, (n=7,199)}} & \multicolumn{3}{c}{\boldmath{\textbf{G}ood \textbf{A}rticles, (n=41,477)}}\\[2pt]
      {} & {Sum} & {Mean}& {STD} & {Sum} & {Mean} & {STD}\\[2pt]
      \hline\rule{0pt}{12pt}$\mathbbm{1}_{unsustain}$	  & 1,151& 0.16& 0.37         & 2,676& 0.06& 0.25    \\[2pt]
        Prom. Time       	  & -- & 2,194.63& 1880.08          & --& 2013.64& 1869.3                             \\[2pt]
        Revisions             & 5,932,445& 824.07& 1434.28               & 20,169,570& 486.28& 1048.59                        \\[2pt]
        Editors	              & 942,461& 130.92& 232.32                   & 3,724,692& 89.8& 176.69                             \\[2pt]
        Comments	           & 93,863& 13.04& 39.18                       & 364,106& 8.78& 26.12                                 \\[2pt]
    \end{tabular}
}
\label{table:data_stats}
\end{table}

To construct the \dataset~population, we look for articles that are currently or have been in the $FA$ or $GA$ status, including those that have been demoted. Wikipedia maintains lists of existing $FA$s and $GA$s \cite{wikipedia2024FA, wikipedia2024GA}. 
In addition, there are Wikipedia pages that list demoted $FA$s, named `Former Featured Articles,' and demoted $GA$s, named `Delisted Good Articles' \cite{wikipedia2024DelistedGA, wikipedia2024FormerFA}. 
We use these four lists to construct the set of articles in \dataset.

\ignore{
    Note that this set of articles contains all relevant articles that have been promoted to $GA$ or $FA$ at least once in their life cycle. However, it does not contain the complete required information regarding promotions and demotions since articles can be promoted/demoted multiple times. For example, the Durian article (see Figure \ref{fig:article_life_cycle}) will be part of the demoted $FA$s list and hence part of the \dataset~population. Note that, it does not appear in either the existing $GA$s or demoted $GA$s lists, as it was demoted from an $FA$ status. However, Durain should also be part of the demoted $GA$s, following the unsustainable articles definition in Section \ref{sec:problem_def}. We address this issue while building the target feature, as described in Section \ref{subsec:target_feature} below.
}

\subsection{Target Feature Construction}
\label{subsec:target_feature}
To construct the target feature per $a \in A$ (i.e.,  $\mathbbm{1}_{unsustain}^{a}$) we extract its promotion to $FA$ and/or $GA$ information. We also extract its demotion information, if any. To do so, we use two complementary methods: article milestones and Wikipedia templates.

\subsubsection{Article milestones}
\label{subsubsec:article_milestones}
Many Wikipedia articles maintain a table on their talk page with information about the article's milestones. Usually, this table contains the dates of promotion, demotion, and reviews that the article has gone through. Figure \ref{fig:article_life_cycle} (bottom) contains a screenshot of the article milestone section of the Durian article.\footnote{Durian talk page: \href{https://tinyurl.com/5f4dtfe7}{https://tinyurl.com/5f4dtfe7}.} We use this information to extract both the promotion/demotion type ($FA$ or $GA$) as well as the date of promotion/demotion.

Relying solely on this information source is deficient. Having an `article milestones' section is not mandatory, and in some cases, the section exists but lacks some of the required information. Therefore, we use a second method to extract promotions/demotions.

\subsubsection{Wikipedia templates}
\label{subsubsec:wikipedia_templates}
In Wikipedia's HTML content, $FA$s and $GA$s are tagged using specific templates. 
The $FA$s and $GA$s are tagged using the \{\{Featured Article\}\} and \{\{Good Article\}\} templates respectively. Once an article is promoted (demoted), the relevant template is added to (removed from) the HTML of the article. Since we process all revisions of each $a \in A$, we can extract the revisions (and dates) in which an article was promoted or demoted.

However, any Wikipedian can add and remove these templates. Edits that do not respect the assessment evaluation process, e.g., due to vandalism or lack of familiarity with Wikipedia guidelines, should be considered when applying this methodology. Hence, we track the changes in the $FA$ and $GA$ templates over time and consider a promotion/demotion \emph{only} if at least one of the following criteria is met:
\begin{itemize}[leftmargin=0.4cm]
    \item The change holds for at least five sequential revisions.
    \item The change holds for at least a month (i.e., 30 days). 
\end{itemize}


Table \ref{table:data_stats} presents basic data statistics for \dataset. In addition to the significant variance observed among articles across various characteristics (e.g., Revisions), we find that the classification problem is heavily imbalanced. Specifically, only 16\% (6\%) of the $FA$s ($GA$s) are categorized as unsustainable.

\subsection{Explanatory Features}
\label{subsec:explanatory_feautres}
We consider a large set of explanatory features for the prediction task. While constructing the different features, we only consider available data before $t_{prom}^a \forall a \in A$. We generate the following six feature sets:

\input{specific_sections/explanatory_features_section}

Overall, we construct 326 explanatory features. The list of all features 
is shared on our project's anonymous GitHub repository.\footnote{The list of features is under the Data folder in our project's 
GitHub repository: 
\href{https://github.com/abrahami/sustainable-high-quality-wikis}{https://github.com/abrahami/sustainable-high-quality-wikis}.}

\section{Experimental Setup}
\label{sec:exper_setup}
We cast the likelihood of an article to be sustainable as a binary classification problem. Per $a \in A$, we aim to predict $\mathbbm{1}_{unsustain}^{a}$ (see Equation \ref{eq:class_indic}). Each article $a$ is represented by an array of explanatory features (detailed in Section \ref{subsec:explanatory_feautres}), extracted from the ``history'' of the article from the time it was created to when it was promoted. That is, our binary classification problem is defined as:$\mathbbm{1}_{unsustain}^{a} \sim Data([t_{birth}^a, t_{prom}^a])$.


\paragraph{Demotion time} While some of the articles in \dataset~have already been demoted some have not (yet). Undemoted articles still have the potential to be demoted in the future and become unsustainable. In practice, we only know partial information, which means we face a \emph{right censored} data problem.

To overcome this challenge, we \emph{exclude} articles that have been recently promoted and did not have ``enough time'' to be considered for demotion. We exclude articles promoted after 2018 (2019) for the $FA$ ($GA$) model, based on the median time of $t_{demo}^a-t_{prom}^a$.

\subsection{Classification Models}
\label{subsec:classification_models}
We experiment with an array of algorithms ranging from simple logistic regression to more sophisticated neural networks. We do not expect complex deep learning and transformer architectures to outperform standard ML models in our settings, since we formulate \dataset~as tabular data \cite{shwartz2022tabular}.\footnote{We do take linguistic and time-related features into account, but formulate those as tabular features.}

We report only the results obtained from the Gradient-Boosted-Decision-Trees (GBT) model \cite{friedman2002stochastic}, which achieved comparable results to other ensemble models such as Random Forest \cite{breiman2001random} and CatBoost \cite{dorogush2018catboost}. We use a deviance loss function for the GBT model. We use the sklearn \cite{pedregosa2011scikit} implementation of the algorithm, with the default parameters of the package. Specifically, we build the model with 100 estimators (i.e., trees). We also tested different approaches to optimizing some of the model parameters.

We execute all algorithms in an ablation manner to learn the importance of the different feature sets. We make the code developed as part of the research and the \dataset~data public in our project's 
GitHub repository.\footnote{The project's 
repository:\\ 
\href{https://github.com/abrahami/sustainable-high-quality-wikis}{https://github.com/abrahami/sustainable-high-quality-wikis}.}

\subsection{Evaluation}
\label{subsec:evaluation}
We report population-level classifier performance using bootstrapped scores across 100 model runs. To analyze \emph{article-level} trends, we perform five-fold cross-validation and use the article-level predictions\footnote{We use the prediction per article from the single fold in which the article is excluded from the training data.} for error analysis and model interpretability. 
We report the average Area under the Receiver-Operator Curve (`AU-ROC'), and the Macro F1 per model. 
Our models for $FA$ and $GA$ are heavily imbalanced, with 16\% and 6\% in the unsustainable class, respectively. Thus, we also record the following over the unsustainable (minority) class: Precision@\{2\%, 5\%\}, Precision, Recall, and F1.


\section{Results and Analysis}
\label{sec:results}

\begin{table}
\centering
\normalsize    
\caption{Modeling results. A compositional model of all features outperforms other alternatives in both $FA$ and $GA$ use cases. We report the average measures ($\pm std$) over 100 resampled bootstrap iterations. We calculate the measures over the positive class (unsustainable articles) besides Macro-F1. The baseline is a model that uses a single feature -- Num-of-Revisions-Normalized.}
{
\begin{tabular}{l@{\Hquad}l@{\Hquad}|c@{\quad}c@{\quad}c@{\Hquad}|c@{}}
      {Type} & {Feature Set} & {Precision} & {Recall} & {F1}& {Macro F1}\\[2pt]
      \hline\rule{0pt}{12pt}
        \multirow{8}{*}{\rotatebox[origin=c]{360}{$FA$}} &
        Baseline                           & 0.31$\pm$0.03& 0.43$\pm$0.14& 0.35$\pm$0.03& 0.58$\pm$0.01\\[2pt]
    \cdashline{2-6} \rule{0pt}{12pt} &
    Network                                 & 0.33$\pm$0.02& 0.39$\pm$0.03& 0.36$\pm$0.02& 0.58$\pm$0.01\\[2pt] &
    Topics                                  & 0.46$\pm$0.04& 0.39$\pm$0.06& 0.42$\pm$0.03& 0.64$\pm$0.01\\[2pt] &
    Team Compos.                            & 0.41$\pm$0.02& 0.50$\pm$0.04& 0.45$\pm$0.02& 0.65$\pm$0.01\\[2pt] &
    Discussions                             & 0.39$\pm$0.02& 0.57$\pm$0.05& 0.46$\pm$0.02& 0.65$\pm$0.01\\[2pt] &
    Edit History                            & 0.44$\pm$0.03& 0.57$\pm$0.06& 0.49$\pm$0.02& 0.67$\pm$0.01\\[2pt] &
    User Experience                         & 0.54$\pm$0.03& 0.62$\pm$0.04& 0.58$\pm$0.02& 0.73$\pm$0.01\\[2pt] &
    All                                     & \textbf{0.62}$\pm$0.03& \textbf{0.68}$\pm$0.04& \textbf{0.65}$\pm$0.02& \textbf{0.78}$\pm$0.01\\[2pt]
    \hline\rule{0pt}{12pt}
    
    \multirow{9}{*}{\rotatebox[origin=c]{360}{$GA$}} &
        Baseline                           & 0.38$\pm$0.06& 0.13$\pm$0.09& 0.18$\pm$0.02& 0.57$\pm$0.01\\[2pt]
    \cdashline{2-6} \rule{0pt}{12pt} &
    Network                                 & 0.19$\pm$0.01& 0.25$\pm$0.02& 0.22$\pm$0.01& 0.56$\pm$0.01\\[2pt] &
    Topics                                  & 0.30$\pm$0.04& 0.25$\pm$0.07& 0.27$\pm$0.03& 0.60$\pm$0.01\\[2pt] &
    Team Compos.                            & 0.26$\pm$0.01& 0.28$\pm$0.02& 0.27$\pm$0.01& 0.60$\pm$0.01\\[2pt] &
    Discussions                             & 0.21$\pm$0.02& 0.42$\pm$0.11& 0.27$\pm$0.03& 0.58$\pm$0.01\\[2pt] &
    Edit History                           & 0.33$\pm$0.02& 0.30$\pm$0.03& 0.31$\pm$0.02& 0.63$\pm$0.01\\[2pt] &
    User Experience                         & 0.46$\pm$0.02& 0.45$\pm$0.02& 0.45$\pm$0.02& 0.70$\pm$0.01\\[2pt] &
    All                                     & \textbf{0.50}$\pm$0.03& \textbf{0.47}$\pm$0.03& \textbf{0.48}$\pm$0.02& \textbf{0.72}$\pm$0.01\\[2pt]
    \end{tabular}
}
\label{table:concise_bootstrap_res}
\end{table}

\begin{figure}
    \centering
    \includegraphics[scale=0.44]{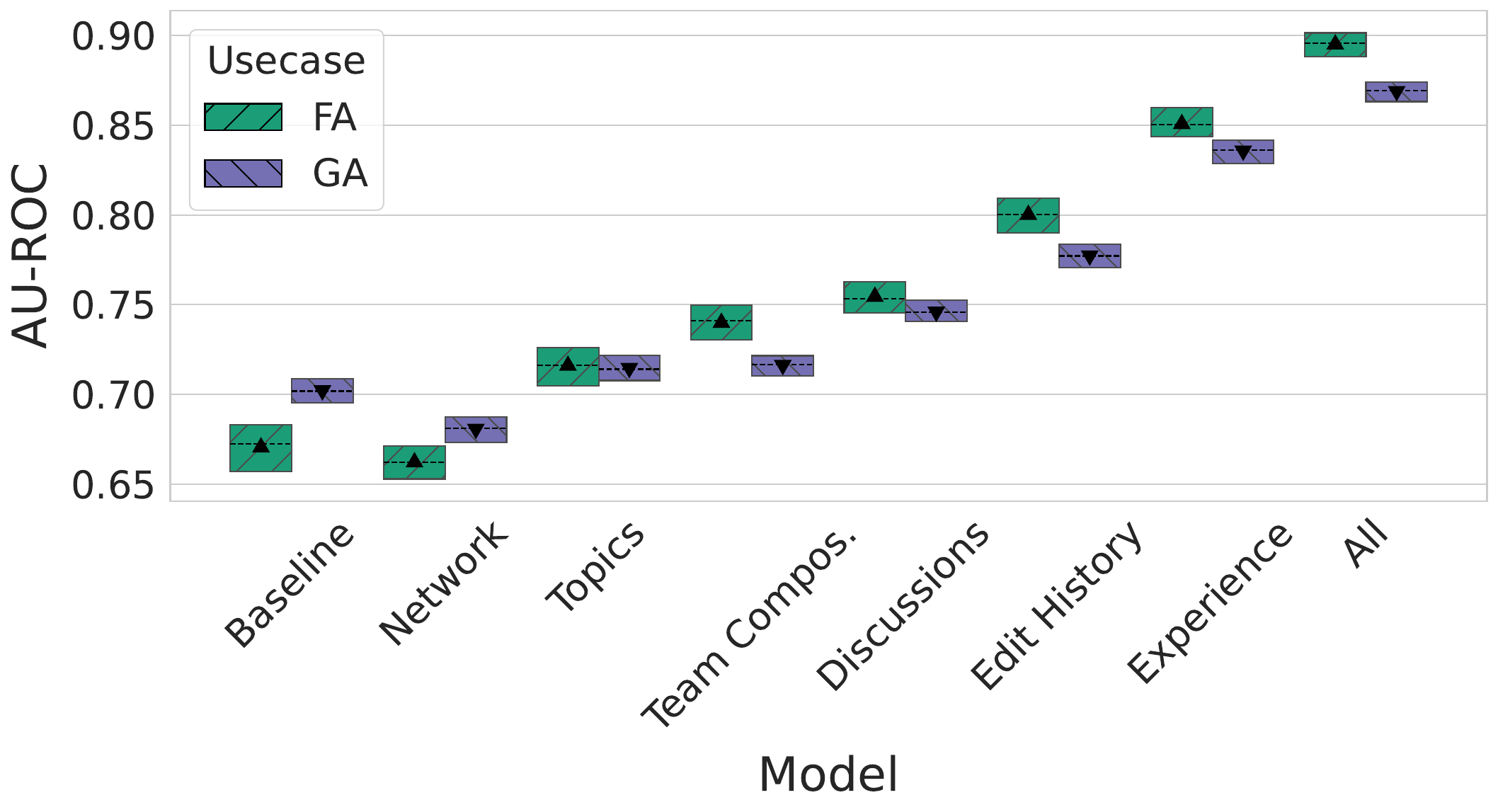}
    \caption{Models performance. We present the aggregate values over 100 resampled bootstrap iterations. Boxplots represent the interquartile range (IQR). Horizontal dashed lines indicate the median values. The black upward (downward) triangle marks the average value of the $FA$ ($GA$) use case. AU-ROC y-axis is the Area under the Receiver-Operator. A random model achieves a 0.5 AU-ROC.}
    \label{fig:models_performace}
\end{figure}

The results obtained by each model in the various ablation settings are detailed in Table \ref{table:concise_bootstrap_res}. We visualize the AU-ROC performance of each feature set in Figure \ref{fig:models_performace}. We also include the full results of both the bootstrapped scores and the five-fold-cv methods in Appendix Tables \ref{table:full_bootstrap_res} and \ref{table:full_5_fold_cv_res}. Except for the `Network' features, using each feature set independently performs well -- yielding significant improvement compared to the baseline model. For both the $FA$ and $GA$ articles, combining all feature sets consistently yields better performance across all evaluation metrics, outperforming any individual set.

Using a single feature type, the highest Macro F1 and AU-ROC are obtained by the User Experience feature set. This holds for both $FA$ and $GA$ use cases.  This highlights the importance of intellectual capital while building a collaborative team \cite{nahapiet1998social, coff1997human}. Specifically, having Wikipedians who are experienced in editing and contributing to successful projects is crucial in creating sustainable articles. We note that while trailing behind, the Edit History feature set alone achieved a decent ($\thicksim15\%$) improvement over the baseline model. On the other hand, the Topics feature set, although comprised of a large set of features, does not outperform other models.

Next, we analyze the data and the best-performing model. We only present results for $FA$s due to space constraints. However, we observe similar patterns for $GA$s.



\subsection{Association Heatmaps}
\label{subsec:association_heatmaps}

\begin{figure*}[t]
\normalsize
    \centering
    \captionsetup[subfigure]{oneside,margin={1.1cm,0cm}}
    \subfloat[{\label{subfig:heatmap_a}}]{{\includegraphics[scale=0.165]{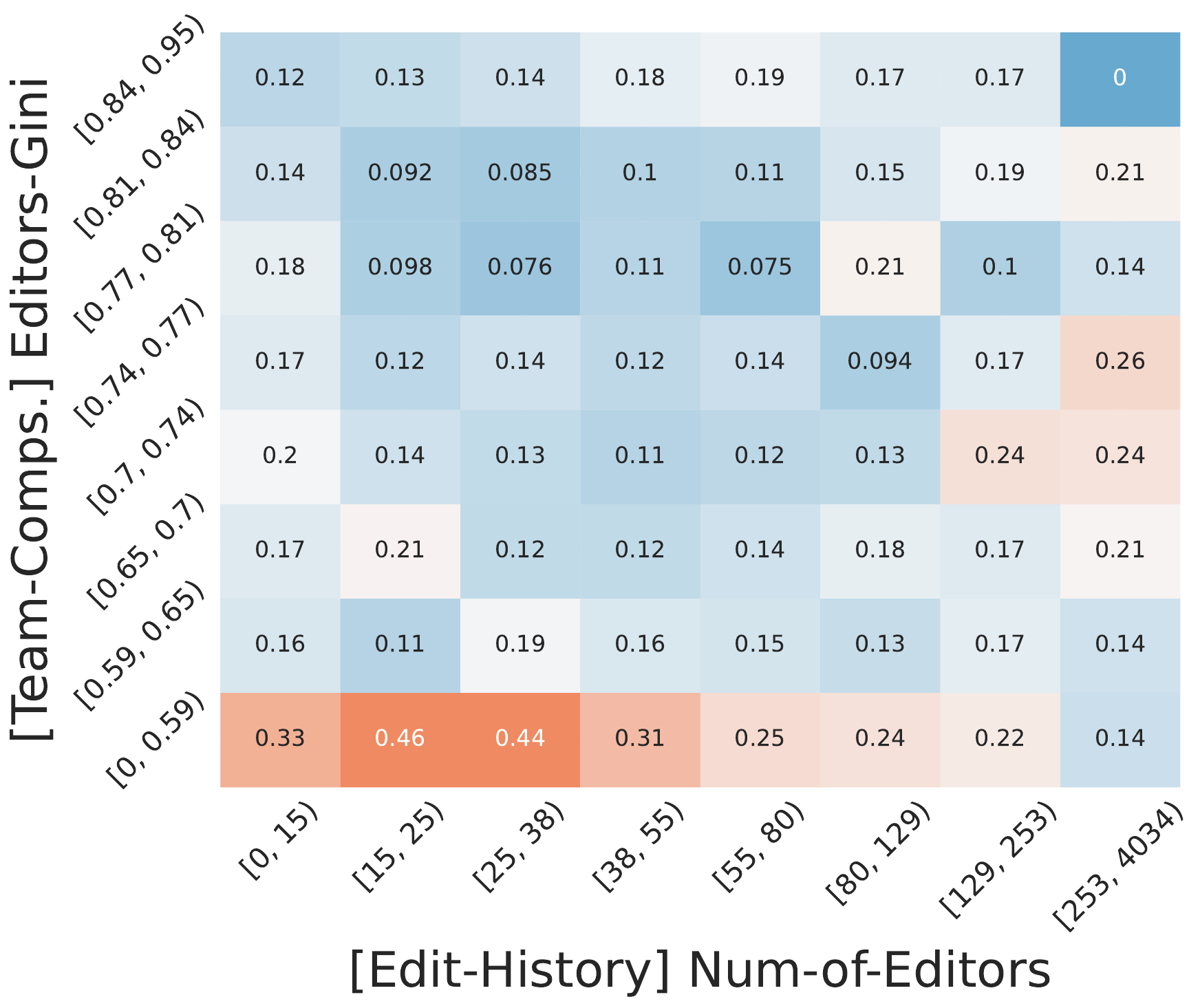}}}
    \quad
    \captionsetup[subfigure]{oneside,margin={1.2cm,0cm}}
    \subfloat[{\label{subfig:heatmap_b}}]{{\includegraphics[scale=0.165]{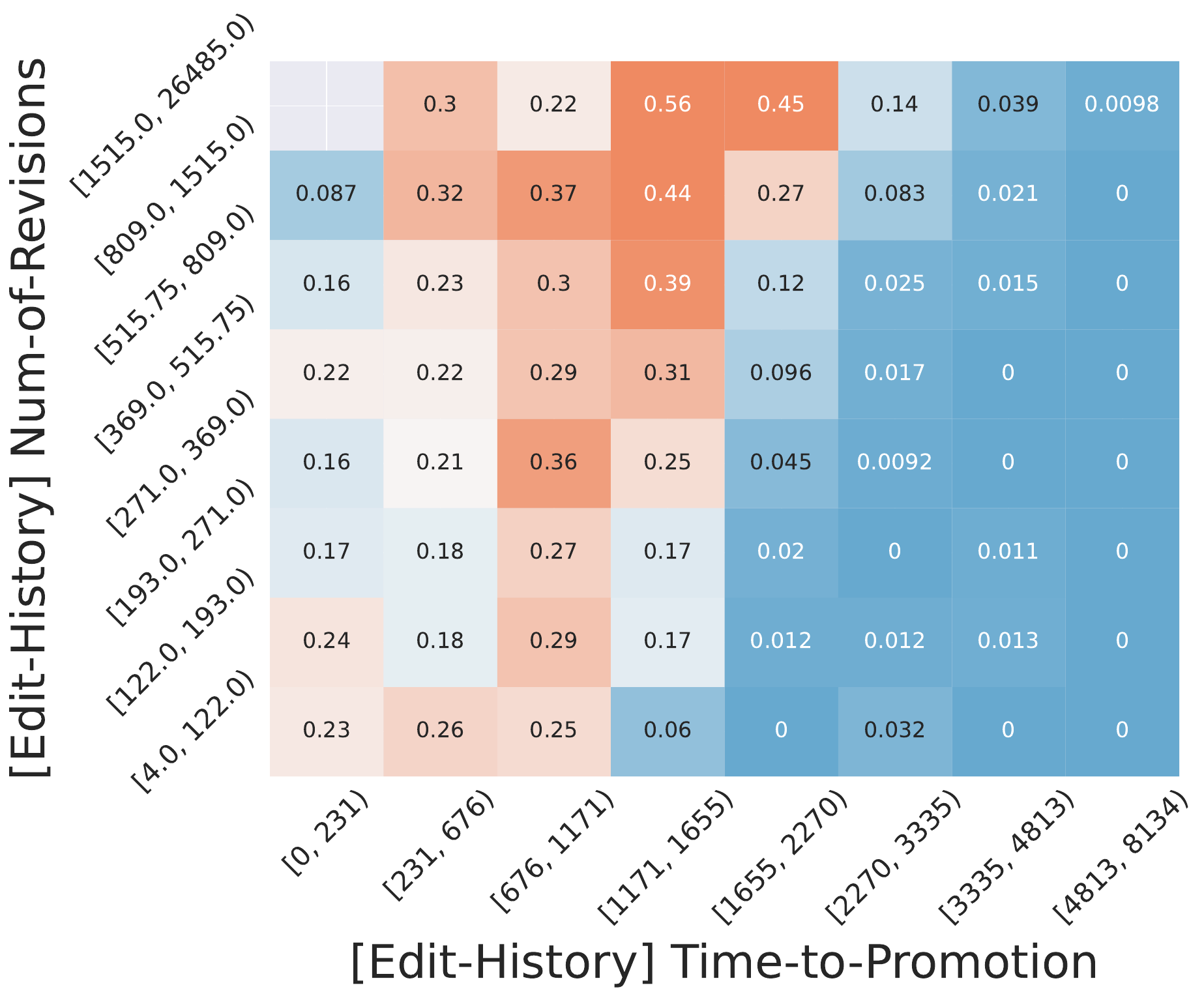}}}
    \quad
    \captionsetup[subfigure]{oneside,margin={0.6cm,0cm}}
    \subfloat[{\label{subfig:heatmap_c}}]{{\includegraphics[scale=0.17]{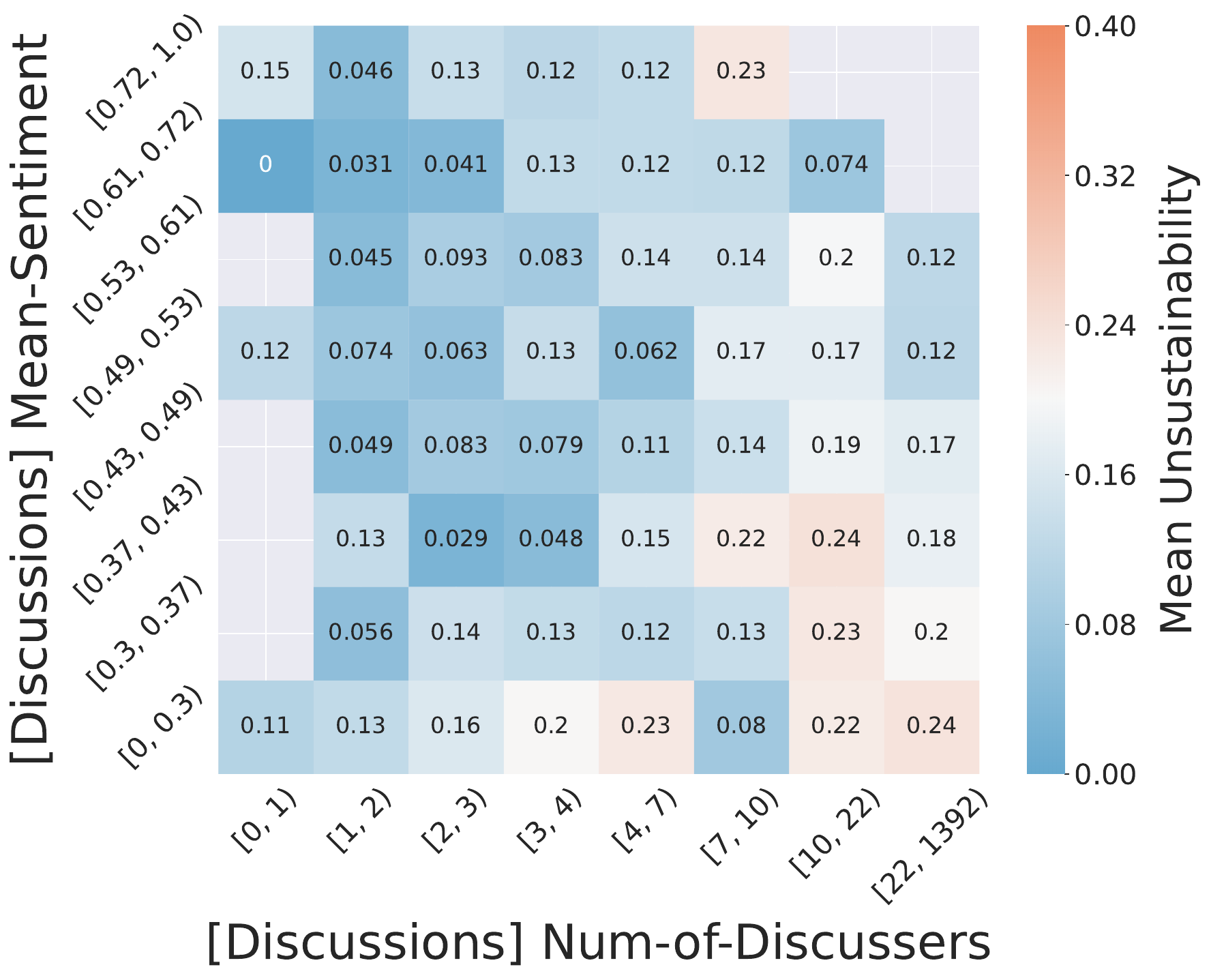}}}

    \caption{Heat maps of feature interactions with unsustainability. Unsustainability is distributed among feature pairs in insightful ways. The value in each cell is the mean unsustainability per bin. We omit presenting values where the number of cases is too small (<10).}
    \label{fig:heatmaps}
\end{figure*}

While the classification model achieves decent results, we still aim to explore how specific feature(s) are associated with (un)sustainability. To do so, we plot the distribution of $\mathbbm{1}_{unsustain}$ over two explanatory features in a heatmap. Figure \ref{fig:heatmaps} highlights three heatmap examples.

The trade-off between the number of contributors and how work is distributed between them has been long questioned \cite{ortega2008inequality,arazy2010determinants}. Figure \ref{subfig:heatmap_a} highlights this balance for our study. It is clear that a lower Gini (unequal contribution) among editors is strongly associated with unsustainability. We also observe that having (too) many collaborators is associated with unsustainability.

Figure \ref{subfig:heatmap_b} demonstrates that the longer it takes for an article to be promoted, the more likely it is to be sustainable. Interestingly, a high number of revisions (y-axis) in a short period of time (x-axis) is strongly associated with unsustainability, suggesting that editing articles slowly over time produces better results that, possibly rushed, high-volume busts of editing.  

Figure \ref{subfig:heatmap_c} highlights that many editors expressing negative sentiment in talk pages put the article at risk of unsustainable status. Few editors expressing positive sentiment in talk pages is more likely to produce sustainable articles. This expands the existing understanding of the relation between discussion patterns and quality assessment in Wikipedia \cite{kittur2008harnessing, stvilia2005information}.

\subsection{Error Analysis}
\label{subsec:error_analysis}
We further conducted an error analysis, focusing on the False Positive (FP) predictions of the best-performing model.
FP predictions are interesting as they can shed light on articles that have not officially been demoted but are possibly at risk. 

To further investigate false positives, articles predicted to be demoted by our model but not currently demoted, we use the Wikipedia reviewing system. As illustrated in the bottom table in Figure \ref{fig:article_life_cycle}, an article can go over multiple review types throughout its life cycle. The Durian article had three distinct review types. 
As we elaborate in Section \ref{subsec:target_feature}, we use these reviews to identify promoted and demoted articles, but note that we did not use data from the review system to construct our prediction features. 

\begin{figure}
    \centering
    \includegraphics[scale=0.35]{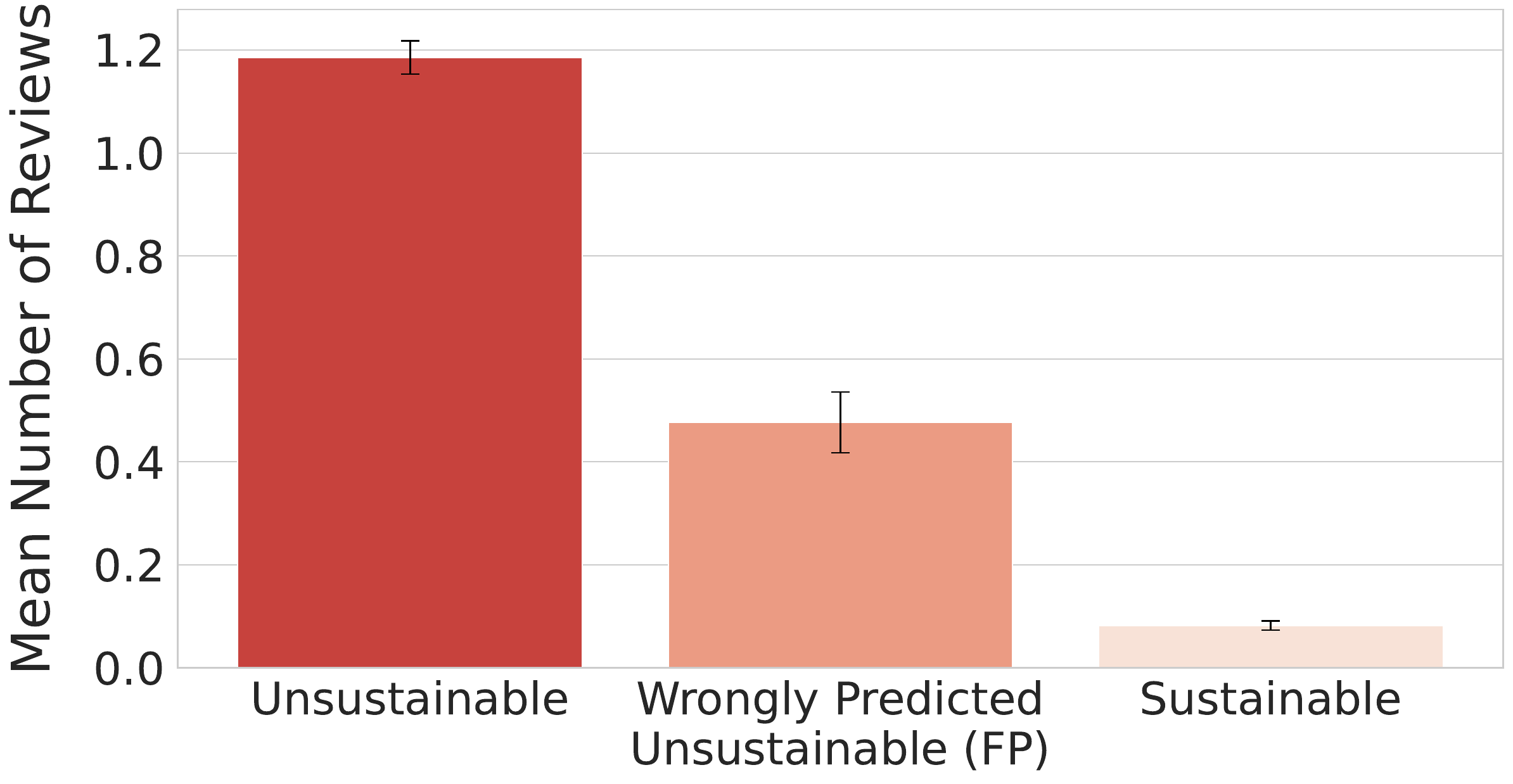}
    \caption{Average number of reviews per population, $FA$ use case. The FP population is reviewed significantly more frequently compared to the sustainable articles population, which highlights their demotion potential -- identified by the Wikipedia community. 
    Naturally, the number of reviews for the unsustainable population is the highest since these demoted articles go over at least a single review.}
    \label{fig:fa_error_analysis_based_reviews}
\end{figure}

Not all $FA$ (or $GA$) reviews lead to a demotion. In many cases, the article keeps its existing status. However, even the fact that an article goes over a review is a signal that the community identified potential issues with the article.\footnote{Most $FA$s and $GA$s have not been reviewed at all since their promotion.} We use the review information to see if FP articles tend to be reviewed more frequently. 

Figure \ref{fig:fa_error_analysis_based_reviews} highlights that our hypothesis truly holds. Wrongly Predicted Unsustainable (FP) articles are being suspected by the Wikipedia community -- reflected by a significantly higher number of reviews compared to the sustainable articles population. Note that unsustainable articles have at least one review since they were demoted at least once. We reiterate that the prediction model is \emph{not exposed} to the review information at all. The model is fed only with information in the [$t_{birth}^a, t_{prom}^a$] timespan. The reviews we analyze here occurred after $t_{prom}^a$.

We conducted the same analysis using the top-100 FP articles. 
As anticipated, we found an even stronger trend, with the average number of reviews for the FP articles exceeding 0.9. This pattern is illustrated in Appendix Figure \ref{fig:erro_analysis_based_reviews_top_100}. 
Additionally, Appendix Table \ref{table:top_10_false_positive} provides a list of the top ten articles from the FP population, and Appendix Table \ref{table:demotion_potential_from_latests_years_promoted} lists the top ten articles with the highest unsustainability likelihood that were recently promoted. These articles have a high probability of being unsustainable according to our model. We expect that these articles are at high risk of demotion.



\subsection{Corpus Size}
\label{corpus_size}
\dataset~is a relatively large corpus. However, 
the number of positive instances is small. We hypothesize that a larger dataset, with more positive examples for the model to learn from, has the potential to improve its overall performance. In this part, we examine (i) The impact of corpus size on the model's performance; (ii) Whether the model's performance is saturated as the corpus size increases.

For that, we split \dataset~into contiguous chunks of articles based on the \emph{promotion year} of each. Using these chunks, we create 14 sequential sub-datasets, $\dataset_{[t]}: t \in [2005, 2018]$. Each $\dataset_{[t]}$ contains all articles promoted by year $t$. Each $\dataset_{[t]}$ contains the same articles that $\dataset_{[t-1]}$ has, as well as the articles that have been promoted on year $t$. $\dataset_{[2018]}$ contains the entire dataset.

We then build and evaluate classifiers for each $\dataset_{[t]}:t \in [2005, 2018]$ independently. Since each $\dataset_{[t]}$ is characterized by a different imbalanced ratio, we evaluate each by the \emph{over-performance} of the model, compared to a random model. We observe that the model's performance has a clear increasing trend and has not saturated as corpus size expands. We see this trend on both the AU-ROC and F1 measures tested. We illustrate this analysis in Appendix Figure \ref{fig:corpus_size_impacts_performance}. This analysis, particularly the continuous upward and unsaturated trend of both curves, emphasizes the critical role of corpus size for our classification task. It also underscores the potential benefits of augmenting \dataset~with future demoted articles and applying the success sustainability metric to larger datasets with more positive data points or less imbalanced.


\subsection{Model Interpretation}
\label{subsec:model_interpretation}
To support RQ2, we use the SHAP framework \cite{lundberg2017unified} to shed light on the specific features that impact the model's prediction. A SHAP value is calculated for each explanatory variable (feature) and input instance (i.e., an article in our case) for a specific model. A high (low) SHAP value indicates the positive (negative) impact of the feature on the prediction of the specific instance. Looking at the aggregate values for a specific feature provides a way to understand feature importance and their positive/negative impact.

\begin{figure}[t]
    \centering
    \includegraphics[scale=0.46]{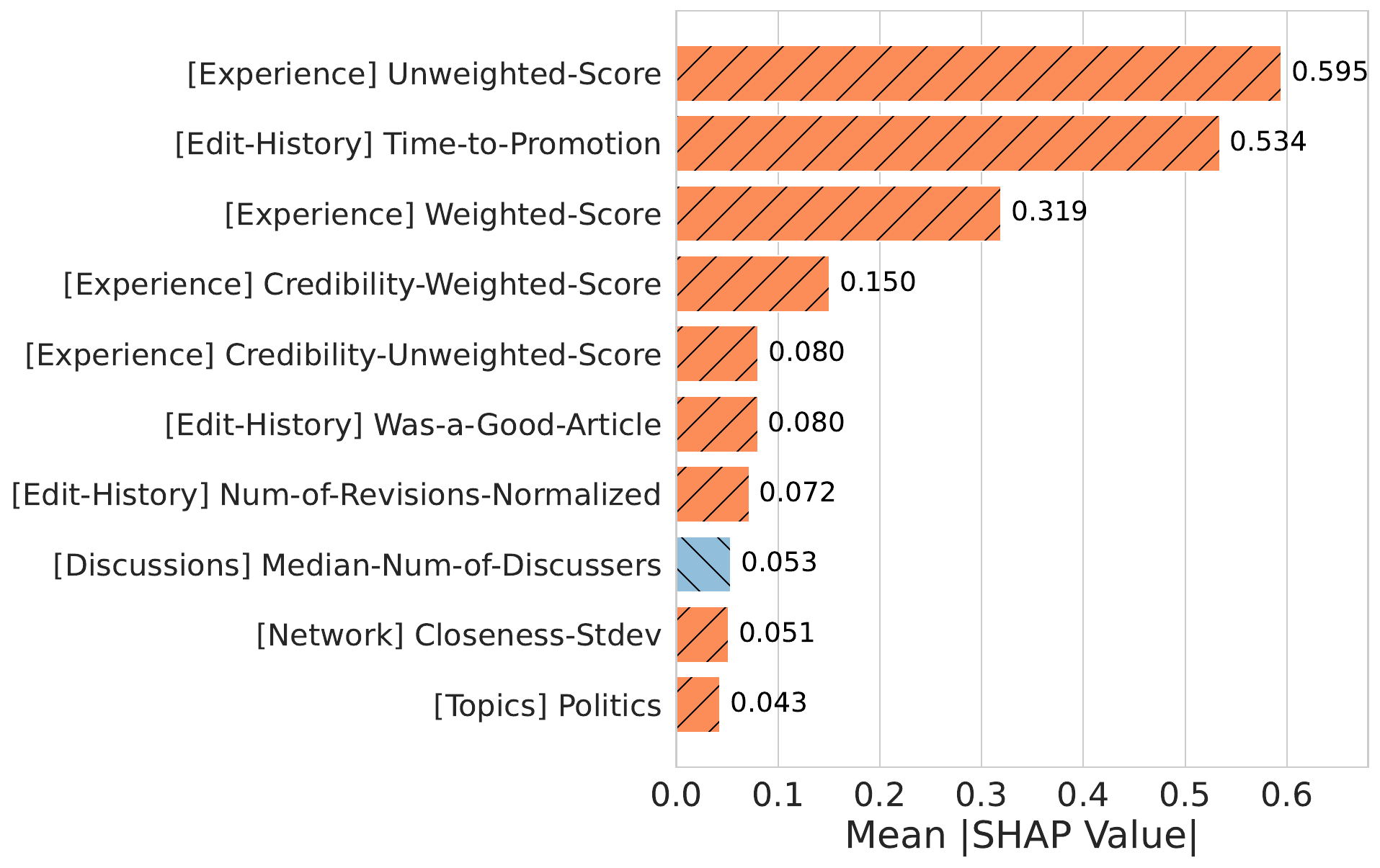}
    \caption{Top ten SHAP features importance. Almost all feature sets are represented in the top ten, while the Experience set dominates the list. Orange (blue) bars correspond to negative (positive) impact on the prediction of unsustainability. }
    \label{fig:shap}
\end{figure}

The absolute SHAP values of the ten most prominent features 
are presented in Figure \ref{fig:shap}.
The bar color and pattern correspond to the positive/negative impact of the feature on the prediction.

\paragraph{Experience is the teacher of all things} Aligned with the results in the first part of this section, the Experience feature set plays the most crucial part in the model. The negative impact (i.e., orange colored bars) reflects that low experience values (i.e., inexperienced team members) are associated with unsustainable articles.

\vspace{-1pt}
\paragraph{Excellence takes time} Three out of the ten dominant features indicate that sustainable success is strongly associated with a long progress that takes time. Time-to-Promotion is simply $t_{prom}^a - t_{birth}^a$ (measured in days) while Num-of-Revisions-Normalized counts the number of revisions an article goes through, divided by Time-to-Promotion. These two indicate that fast-promoted articles, in terms of time and number of edits, are more likely to fail in the future (i.e., be unsustainable). Was-a-Good-Article (see Section \ref{subsec:explanatory_feautres}, feature $EH_5$) points out that articles which \emph{did not} go through a two-step promotion are more likely to be unsustainable, showing again that a slower trajectory to $FA$ with an initial stop at the $GA$ status is more likely to lead to a sustainable article in $FA$. 29.2\% of the articles that were directly promoted to $FA$ are found to be unsustainable, compared to only 9.9\% of those that were first promoted to $GA$.  

\paragraph{Marginal but critical} Each of the three last features comes from a different feature set. Median-Num-of-Discussers highlights that (too) many Wikipedians participating in the talk page discussions increase the likelihood of an article being unsustainable. 
This observation is consistent with previous research on crowd collaboration success \cite{brooks1995mythical, kittur2008harnessing}, which emphasizes that an increase in the number of contributors does not inherently improve the quality of the collaboration but rather requires different coordination strategies.
We assume that the high association of the Politics topic with unsustainability is due to (i) Many controversial issues raised about political items \cite{rad2012identifying}, and (ii) The frequent information updates and new events in the political domain, leading articles becoming quickly outdated and thus being more difficult to sustain.

While not all the dominant features we identified are actionable, many are. Notably, a gradual progression toward attaining recognized high-quality status was found to be strongly associated with sustainable success. This finding suggests implementing a policy of incremental quality assessments, rather than immediate promotion to the highest assessment level, may be beneficial. In the context of Wikipedia, this approach is exemplified by initially designating articles as $GA$ before considering them for an $FA$ status. Additionally, our analysis highlights the critical role of experienced collaborators with a demonstrated track record of success. This actionable feature should motivate collaborative teams to engage with experienced users and convince them to join their collaborative effort to enhance the likelihood of achieving sustainable success.

\section{Discussion and Limitations}
\label{sec:discussion_limitations}
\paragraph{Other studies of collaborative success} The objective of most existing studies is `Success Prediction.' That is, to research what leads crowd collaboration to initial success. Our objective is different. We aim to measure these collaborations through the lens of whether success holds over time, using our new `Sustainable Success' measure. We see these two objectives as different by nature. However, highlighting the similar (and different) dominant features between `Success Prediction' and `Sustainability Prediction' is insightful.

A key feature that we find to be associated with `Sustainability Prediction' is a prolonged and gradual path toward attaining a recognized high-quality status. This feature is unique to our study and has yet to be recognized in other `Success Prediction' studies.\footnote{Naturally, `Success Prediction' studies are limited in the way this feature can be constructed as their target feature is different than ours.} The participation of experienced team members with a proven history of involvement in successful collaborations, which we find a significant feature, was also found to be significant in a few `Success Prediction' studies \cite{stein2007does,betancourt2016mining}.

For other features, such as number-of-editors, editors-gini, and most network features, while we did not find those to be the most dominant, other `Success Prediction' studies found them as significant predictors \cite{arazy2010determinants,qin2012assessing,raman2020classifying,lin2020wisdom}.

Overall, our findings emphasize the unique features critical to sustainable success, while also identifying shared elements with existing research on the general task of success prediction.

\paragraph{The broader picture} We use Wikipedia as a testbed to answer our research questions. However, quantifying the sustainable success of crowd collaborations goes beyond Wikipedia articles and extends to other online and offline collaborative efforts such as crowdsourced open-source software and entrepreneurial teams \cite{lechler2001social, lazar2020entrepreneurial, daniel2018quality, saadat2020analyzing}. These crowd-collaboration projects span multiple domains and play a critical role in large-scale projects.

In such collaborations, success is not merely determined by the short-term success or the initial hype they make but rather by the longevity and lasting impact they generate. The social dynamics driving these collaborations share parallels with the challenges faced in maintaining the long-term quality and relevance of Wikipedia articles.
We see the importance of our study also in introducing a framework to measure and model the sustainable success of online/offline human crowd collaborations beyond Wikipedia.

\paragraph{Identifying demotion}
Creating \dataset~ required merging multiple sources of data. For example, our initial target feature, 
$\mathbbm{1}_{unsustain}^{a}$, was solely based on the Wikipedia templates data (see Section \ref{subsubsec:wikipedia_templates}). After analyzing the promoted and demoted articles using this data source, we identified many unreliable cases. In most cases, the unreliability is due to unauthorized removal/addition of templates -- edit wars or vandalism. Hence, we decided to modify our approach in three ways: (i) We add restricted conditions on when to use the Wikipedia templates (see Section \ref{subsubsec:wikipedia_templates}); (ii) We define \dataset~based on Wikipedia promoted/demoted official lists (see Section \ref{subsec:dataset_population}); and (iii) We use Wikipedia Timeline as an extra information source (see Section \ref{subsubsec:article_milestones}). We find our modified approach to constructing the target feature more reliable and robust.

\paragraph{Ethical use of data}
In this study, we utilize data from Wikipedia, a platform that provides full access to its data under an open license. Additionally, we adhere to ethical standards, ensuring that Wikipedian identities are not disclosed or analyzed in our models. Our focus remains solely on the content and activity patterns rather than individual contributors, further ensuring the privacy and integrity of Wikipedia's collaborative environment.

\paragraph{Limitations}
This study focuses exclusively on English Wikipedia articles, limiting its scope to non-English projects. Additionally, the data used is restricted to the point in time when the study was conducted, meaning that articles may be demoted in the future, which we cannot account for. As described in Section \ref{sec:data}, our analysis is based solely on Wikipedia's internal dynamics, without incorporating interactions that may occur outside of the platform (e.g., social media).

Finally, our research is based on ``naturally occurring'' data and events, rather than a ``designed experiment''. Since we do not control the available data or its scale, this imposes limits on our ability to make causal claims---particularly in drawing causal conclusions between explanatory factors and sustainable success. Future research that incorporates controlled experiments could provide deeper insights into the causal relationships between key factors and sustainable success, allowing for a more rigorous evaluation of their long-term impact.

\section{Summary}
\label{sec:summary}
Successful collaboratively constructed products can fall in disrepute over time and, in this study, we introduce a novel way to measure crowd collaboration success through the lens of whether success holds over time, which we call ``Sustainable Success.'' Using the \dataset~dataset, a large, longitudinal corpus of Wikipedia articles and metadata, we train models to show that sustainable success---and eventual failure---is predictable. 
Through our analyses of errors and model interpretability, we shed light on the factors that make crowd collaboration sustainable. Among these factors are a prolonged and gradual path toward attaining a recognized high-quality status, and the participation of experienced team members with a proven history of involvement in successful collaborations.


Future research takes two trajectories: (i) Expanding the research to non-English Wikipedia projects, especially as these relate to different cultural and curational practices, 
and (ii) Studying sustained success in other collaborative platforms such as GitHub. 

\section*{Acknowledgments}

This work was supported by the National Science Foundation under Grants No. IIS-2007251 and IIS-2143529.

\bibliographystyle{plainnat}
\bibliography{wiki_umich_www25}


\appendix
\input{specific_sections/appendix}

\end{document}

%% file: specific_sections/explanatory_features_section.tex
\ignore{
    \paragraph{Edit history features} This set of features captures the editing dynamics of the article. We extract and calculate each of these features from the article's revisions history in the [$t_{birth}^a$, $t_{prom}^a$] timespan. This set consists of six features. Figures \ref{subfig:heatmap_a} and \ref{subfig:heatmap_b} present the distribution of four of those features and their association with sustainable success. In Figure \ref{subfig:heatmap_a}, a lower Gini Index among editors is shown to be strongly associated with higher unsustainable success. Figure \ref{subfig:heatmap_b} demonstrates a high number of revisions (y-axis), occurring within a short period of time (x-axis) also correlates positively with unsustainable success. 
}

\paragraph{Edit History features} This feature set, named $EH$, captures the editing dynamics of the article, which have been shown to be associated with performance in Wikipedia \cite{ruprechter2020relating, kittur2008harnessing}. We extract and calculate $EH$ from the article's revisions history in the [$t_{birth}^a$, $t_{prom}^a$] timespan. It includes, number-of-editors ($EH_1$), number-of-revisions ($EH_2$), time-to-promotion ($EH_3$), and reverted-revisions-percentage ($EH_4$) which is the number of reverts out of $EH_2$. We also create normalized features for the first two metrics to account for edit activity and time: $EH_1 / EH_2$ and $EH_2/EH_3$, respectively. 
The Was-a-good-article ($EH_5$) binary feature indicates whether the article was first promoted to $GA$ before a second promotion to $FA$ (see example in Figure \ref{fig:article_life_cycle}). It is only relevant for $FA$s.

\ignore{
    \paragraph{Network features} We refer to network features as the ones extracted from the edit-network $G_a=(V_a, E_a)$ that we construct per $a \in A$. $G_a$ is based on the edit interactions among collaborators. $V_a$ (nodes) and $E_a$ (edges) represent Wikipedians and their interactions, respectively. Numerous scholars show the effect of network structure among collaborators and success \cite{hinds2008social,romero2015coordination}. In Wikipedia, a few studies focus on the network structures and their impact on Wikipedia articles \cite{brandes2009network, jurgens2012temporal, lerner2019network,qin2015influence}. Inspired by \citet{platt2018network} and \citet{lerner2019network}, we build $G_a$ according to the sequential revision history. A directed edge $e \in E_a$ is assigned between two Wikipedians ${v_1, v_2} \in V_a$ if $v_2$ made a revision to article $a$, followed by a revision of $v_1$. Using $G_a$, we extract 27 structural features per article, such as the network density, the average centrality, etc.
}

\paragraph{Network features} Numerous scholars show the effect of network structure among collaborators and success \cite{hinds2008social,romero2015coordination, romero2016social}. In Wikipedia, a few studies focus on the network structures and their impact on Wikipedia articles \cite{brandes2009network, jurgens2012temporal, lerner2019network,qin2015influence}. We refer to network features as the ones extracted from the edit-network $G_a=(V_a, E_a)$ that we construct for each $a \in A$. $G_a$ is based on the edit interactions among collaborators. $V_a$ (nodes) and $E_a$ (edges) represent Wikipedians and their interactions, respectively.\footnote{Anonymous editors are excluded.} Inspired by \citet{platt2018network} and \citet{lerner2019network}, we build $G_a$ according to the sequential revision history. A directed edge $e \in E_a$ is assigned between 
${(v_1, v_2)} \in V_a$ if $v_2$ revised article $a$, directly after a revision by $v_1$. 

We extract the following for each $G_a$: $|V_a|$, $|E_a|$, number-of-triangles, density, number-of-connected-components (regular and strongly), is-biconnected, and number-of-nodes-to-cut. The latter two are calculated based on the largest component of $G_a$ and the last represents the smallest number of nodes that must be removed to cut $G_a$ into two connected components. In addition, we calculate the following metrics for each $v \in V_a$: in-degree, out-degree, centrality, betweenness, and closeness. For each metric, we calculate the average, median, and standard deviation over $V_a$ to have aggregated features per article. 

\ignore{
    \paragraph{Topic features} As we explain in Section \ref{sec:problem_def}, WikiProjects play an essential role in the Wikipedia ecosystem -- both in the content assessment process and the way information is organized. Most Wikipedia articles are associated with multiple WikiProjects. The importance of the association between an article and its WikiProject spans from `Top' to `Low' with two additional categories of 'High' and 'Mid.' For example, the Durian article is associated with twelve WikiProjects. Among them are the `Plants', `Malaysia', and `South Asia' WikiProjects in which Durian is associated with a `High,' `Mid,' and `Low' importance, respectively. We use this information to create topical features per article on a Likert scale (0-4).\footnote{A zero value is assigned when an article is not associated with a WikiProject. A value of four is assigned when an article is associated with a WikiProject as `Top-Importance.'} Since there are over 2,000 English WikiProjects\footnote{Source: \href{https://tinyurl.com/5xyahfm7}{https://tinyurl.com/5xyahfm7}.}, we only process the top 250 most frequent ones.\footnote{Three of the 250 topics were manually removed as they were not purely topical. For example, the `Spoken Wikipedia': \href{https://tinyurl.com/33h98n2p}{https://tinyurl.com/33h98n2p}.}
}
\paragraph{Topic features} Existing scholars highlight the relation between Wikipedia topics to their edit frequency and edit wars \cite{rad2012identifying, wilson2015content}. We hypothesize that incorporating topical features will improve the model's prediction. These features not only enhance the model's accuracy but also decrease the likelihood that the associations we find between other features and sustainability are not confounded by the topic of the article. 

As we note in Section \ref{sec:problem_def}, most Wikipedia articles are associated with multiple WikiProjects. Each WikiProject rates the importance of the article to the WikiProject as `Low,' `Mid,' `High,' or `Top' \cite{wikipediaEngine}. 
We utilize this information to generate topical features for $a \in A$, using a Likert scale ranging from 0 to 4.\footnote{A zero value is assigned when an article is not associated with a WikiProject. A value of four is assigned when an article is associated with a WikiProject as `Top-Importance.'} There are over 2,000 English WikiProjects and many of them are very small \cite{wikipedia2024WikiProjects}. We only use the top 250 by number of articles.\footnote{Three of the 250 topics were manually removed as they were not purely topical. For example, the `Spoken Wikipedia:' \href{https://tinyurl.com/33h98n2p}{https://tinyurl.com/33h98n2p}.} 97.7\% of the articles in the \dataset~dataset belong to at least one of these 250 WikiProjects.

\ignore{
    \paragraph{Team composition features} The way a team of Wikipedians engages with one another and with the edited article is crucial for the team's success and resilience \cite{ruprechter2020relating}. We hypothesize that this is also critical to the success of an article in terms of sustainability. Hence, we construct five team composition features such as \textit{Editors\_Gini}, which represent the Gini inequality coefficient \cite{gini1912variabilita} among the Wikipedians editors of an article. This feature set also includes features that capture the overlap proportion between Wikipedians who edit the article and those who participate in the discussions (on the article's talk page).
}

\paragraph{Experience features} We hypothesize that the experience of an article's editors positively impacts its sustainable success. To construct `Experience' features, we extract the set of editors,${V^a}$, of each article $a \in A$ during [$t_{birth}^a$, $t_{prom}^a$]. We then find the number of edits done by ${V^a}$ in the rest of \dataset, \emph{excluding} article $a$. Note that this measures experience in editing other articles that were eventually promoted. We construct aggregated measures per article $a$ by either (i) Summing all edits or (ii) Weighing the number of edits by the proportional contribution of each $v \in {V^a}$.

We construct two modified features named `Credibility' that consider whether the other articles edited by authors in ${V^a}$ are sustainable articles or not. In this setup, edits to sustainable (unsustainable) articles positively (negatively) contribute to the aggregated measure. We only count sustainable (unsustainable) cases if the promotion (demotion) is known by $t_{prom}^a$. As before, aggregation over ${V^a}$ is done by either (i) Summing all edits; or (ii) Weighing the edits by the proportional contribution of each $v \in {V^a}$.

\ignore{
    \paragraph{Discussions features} The way Wikipedians manage their conversations has been shown to be correlated with various success measures \cite{maki2017roles, stvilia2005information}. We hypothesize that polite and respectful discussions would be positively associated with sustainable success.
    
    We collected all discussions between Wikipedians on the talk page of each article and constructed two types of discussion features. The first type are \emph{linguistic} features. We use pre-trained deep-learning models to predict those. Specifically, we predict the sentiment \cite{hartmann2023more}, formality \cite{babakov2023don}, politeness \cite{danescu-niculescu-mizil-etal-2013-computational,wang-jurgens-2018-going}, toxicity \cite{logacheva2022paradetox}, and certainty \cite{pei2021measuring}~per comment in a discussion. We then aggregate those over all comments per article to have, for example, the average sentiment.
    
    The other type of features are related to the \emph{structure} of the discussions. Among those, for example, are the average number of responded comments, the number of discussers, and the Gini index \cite{gini1912variabilita} calculated based on the contribution of all discussers. We extract 20 and 16 features of the first and second types, respectively. Figure \ref{subfig:heatmap_c} highlights that many discussers expressing negative sentiment positively correlated with unsustainability. Few discussers who express positive sentiment are more likely to produce sustainable articles.
}

\paragraph{Discussions features} The way Wikipedians handle their discussions has been shown to be correlated with various success measures \cite{maki2017roles, stvilia2005information}. We expect to find a significant correlation also with sustainable success. For example, we assume that polite and respectful discussions would be positively associated with sustainable success since these types of discussions may be more likely to incorporate diverse perspectives.
    
We collected \emph{all} historical discussions between Wikipedians (i.e., discussers) on the talk page of each article and constructed two types of discussion features. The first type are \emph{linguistic} features. We use pre-trained deep-learning models to measure those. Specifically, we measure the sentiment \cite{hartmann2023more}, formality \cite{babakov2023don}, politeness \cite{danescu-niculescu-mizil-etal-2013-computational,wang-jurgens-2018-going}, toxicity \cite{logacheva2022paradetox}, and certainty \cite{pei2021measuring}~for each comment in a discussion. We then aggregate each, using the average, median, and percentage of high values.\footnote{We use previously identified thresholds to define high values using each model} To account for uneven contributions across the discussers of talk pages that might bias our measures, we also calculate the mean and median values for each discusser and then aggregate these values across all discussers.

The other type of features are related to the \emph{structure} of the discussions. We list these 15 features in Appendix Table \ref{table:discussions_structural_features}. Among those features are number-of-discussers, mean-responded-comments, and the discussers-gini \cite{gini1912variabilita}, which is a measure of statistical dispersion commonly used to measure inequality. A Gini index of 0.0 indicates perfect equality where all editors contribute the same amount of content (i.e., comments in the case of discussions), while 1.0 indicates perfect inequality where a single individual generates all content. 

\ignore{
    \paragraph{Team composition features} The way a team of Wikipedians engages with one another and with the edited article is crucial for the team's success and resilience \cite{ruprechter2020relating}. We hypothesize that this is also critical to the success of an article in terms of sustainability. Hence, we construct five team composition features such as \textit{Editors\_Gini}, which represent the Gini inequality coefficient \cite{gini1912variabilita} among the Wikipedians editors of an article. This feature set also includes features that capture the overlap proportion between Wikipedians who edit the article and those who participate in the discussions (on the article's talk page).
}

\paragraph{Team Composition features} The way a team of Wikipedians engages with one another and with the edited article is crucial for the team's success and resilience \cite{ruprechter2020relating}. We hypothesize that this is also correlated with sustainable success. This feature set, named $TC$, includes editors-gini ($TC_1$) and ip-based-edits-percentage ($TC_2$). $TC_1$ corresponds to the Gini inequality coefficient \cite{gini1912variabilita} among the Wikipedian editors. It is calculated in the same way as for discussers-gini (see previous paragraph) while taking into account article edit distribution among Wikipedians. $TC_2$ represents the fraction of revisions done by anonymous editors.

$TC$ also includes features representing the relation between two populations: editors of the main article ($V^a)$ and discussers of the talk page ($D^a)$. We create the following: fraction of main article editors who also continue to the talk page discussion ($TC_3$), fraction of editors of the talk page who also contribute to the main article ($TC_4$), and distribution-diff ($TC_5$). $TC_5$ captures how different is the contribution between editing and discussing among Wikipedians. The contribution is calculated as the proportional edits (or comments) over all revisions (or discussions). We first calculate the contribution difference between edits and discussions per $ a \in {V^a \cup D^a}$ and then average the differences.  

%% file: specific_sections/appendix.tex
\newpage
\section{Appendix}
\label{sec:appendix}

\begin{figure}[ht]
    \centering
    \includegraphics[scale=0.42]{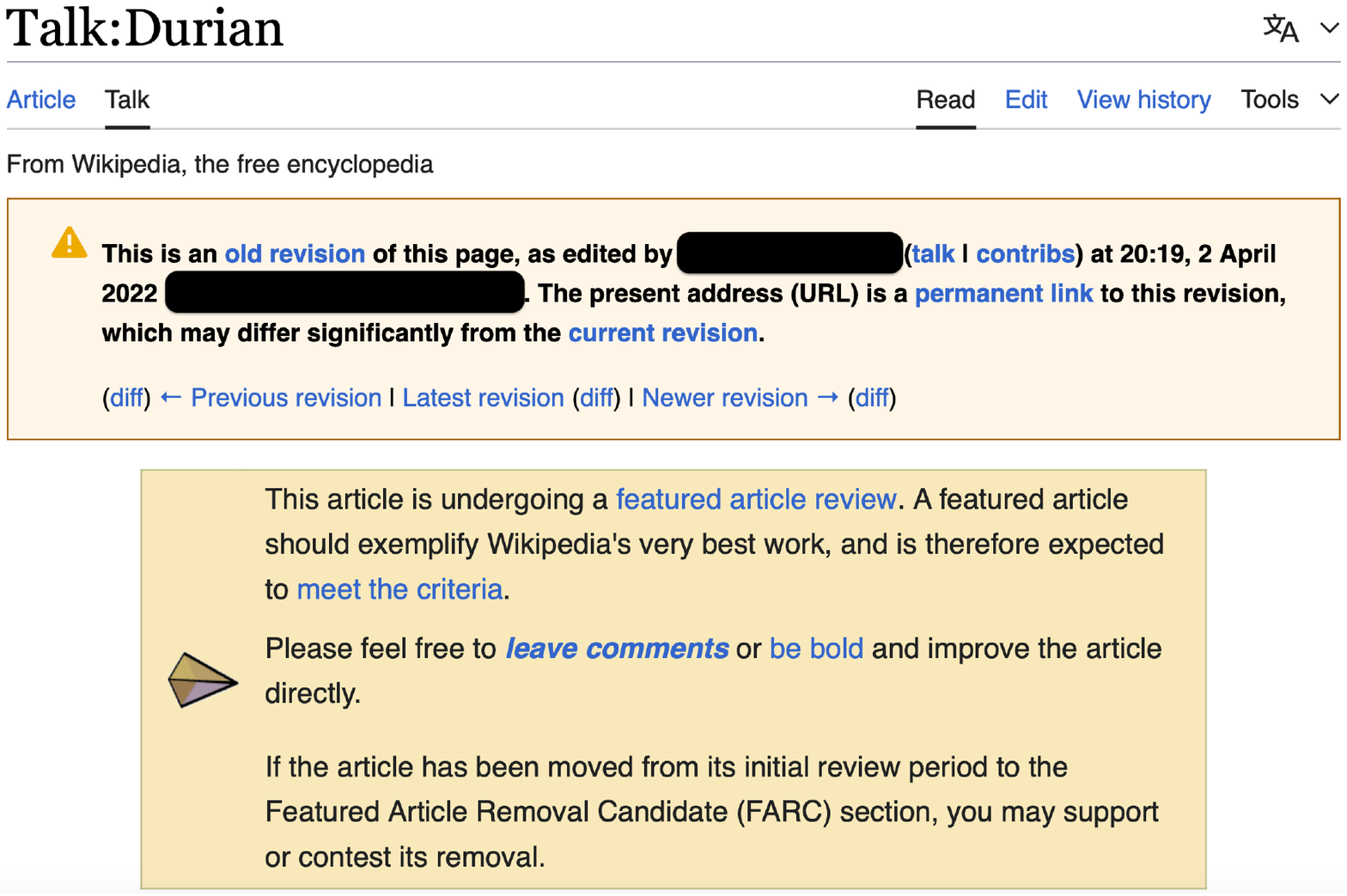}
    \caption{Review process alert of the Durian article. The screenshot was taken from an April 2022 version of the talk page. Unfortunately, the article was demoted in June 2022 after this review process. We conceal all identities to preserve editors' right to remain anonymous.}
    \label{fig:durian_review_alert}
\end{figure}

\begin{figure}[ht]
    \centering
    \includegraphics[scale=0.4]{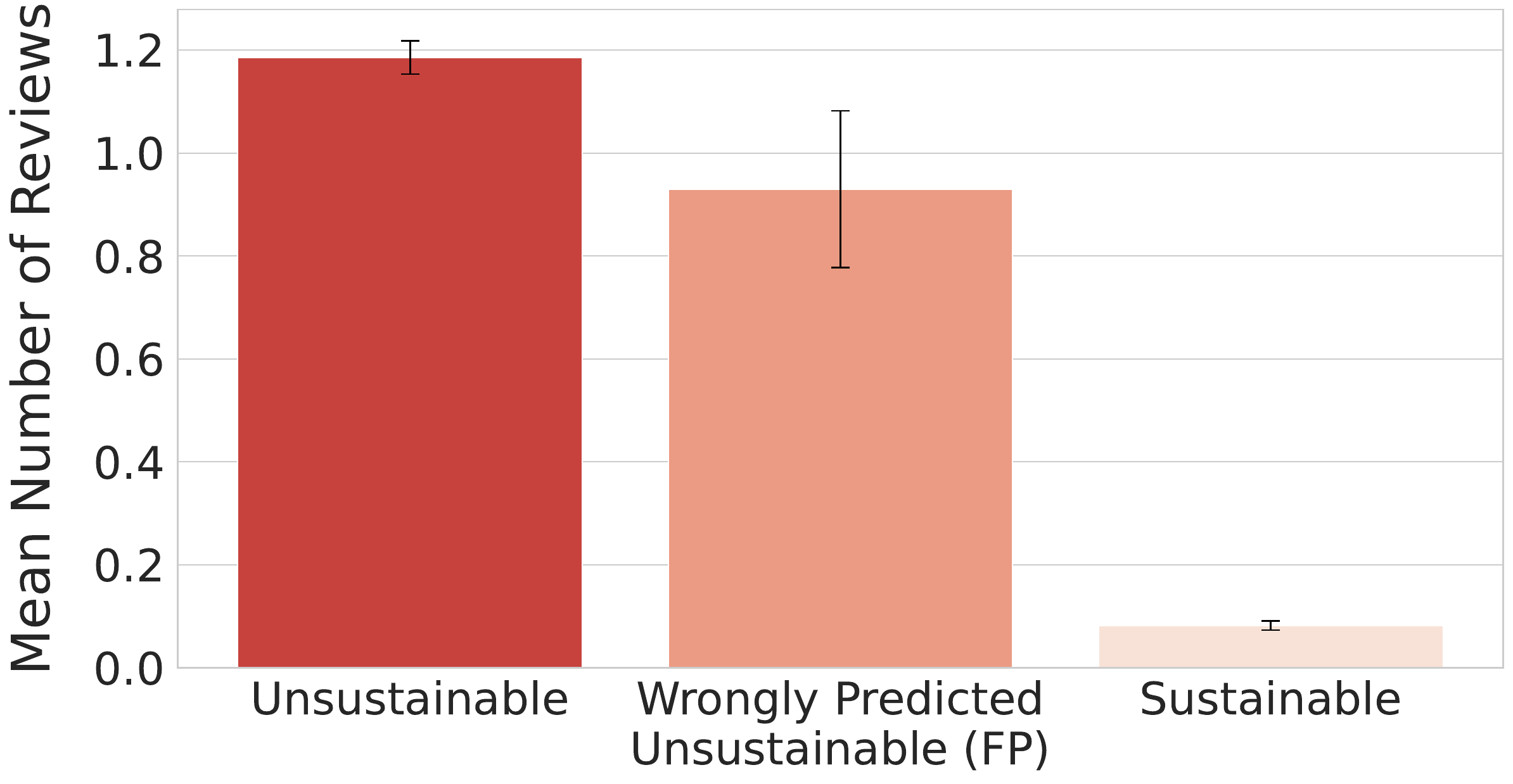}
    \caption{Average number of reviews per population, $FA$ use case. The FP population includes only the top-100 articles predicted by the model to be unsustainable (i.e., articles are sorted by the predicted likelihood of the model). The FP population is reviewed significantly more frequently compared to the sustainable articles population, which highlights their demotion potential identified by the Wikipedia community. This is aligned with the pattern we illustrate in Figure \ref{fig:fa_error_analysis_based_reviews} when taking the whole FP population (and not only the top-100 articles). Naturally, the number of reviews for the unsustainable population is the highest since these articles go over at least a single review before being demoted.}
    \label{fig:erro_analysis_based_reviews_top_100}
\end{figure}

\begin{table*}
\centering
\footnotesize
\caption{Discussion features related to the structure of discussions. Details about this feature set are provided in Section \ref{subsec:explanatory_feautres}.}
{
    \begin{tabular}{p{4.3cm}p{11.3cm}p{0.6cm}} 
     \multicolumn{1}{c}{Feature Name} &
     \multicolumn{1}{c}{Short Description} & 
     \multicolumn{1}{l}{Type}\\[2pt]
      \hline\rule{0pt}{6pt}Num-of-Discussions       & The number of discussion threads on the talk page. &               Int    \\[2pt]
        Num-of-Comments                             & The number of comments over all discussion threads &               Int    \\[2pt]
        Num-of-Discussers                           & The number of Wikipedian discussers over all discussion threads, not including anonymous ones. & Int  \\[2pt]
        Discussers-Gini                             & Gini index (inequality metric) among  Wikipedian discussers. &     Float    \\[2pt]
        Mean-Discussers                             & The average number of Wikipedian discussers among discussions.&    Float    \\[2pt]
        Median-Discussers                           & The median number of Wikipedian discussers among discussions.&     Float    \\[2pt]
        Mixed-Discussers-Comments                   & The proportion of comments that had two discussers editors, which indicates disagreements.  & Float\\[2pt]
        Direct-Discusser-Interactions               & The number of discussed pairs that had direct interaction (commented each other).& Int\\[2pt]
        Indirect-Discusser-Interactions             & The number of discussed pairs that had direct or indirect interaction (commented the same thread). & Float\\[2pt]
        Mean-Triangles-Direct-Interactions          & The average number of triads between Wikipedian pairs that had direct interaction. & Float\\[2pt]
        Mean-Triangles-Indirect-Interactions        & The average number of triads between Wikipedian pairs that had direct or indirect interaction. & Float\\[2pt]
        Mean-Depth                                  & The average depth of discussion threads over all discussions. & Float \\[2pt]
        Mean-Comments                               & The average number of comments per discussion.&                   Float\\[2pt]
        Mean-Responded-Comments                     & The proportion of comments being replied. This is calculated only over the zero-level comments (i.e., not including comments to comments).& Float \\[2pt]
        Time-to-Reply                               & The average time, measured in seconds, it takes to respond to a comment. This is calculated only over the zero-level comments (i.e., not including comments to comments).& Float \\[2pt]      
    \hline
    \end{tabular}
\label{table:discussions_structural_features}
}
\end{table*}

\begin{table*}
\centering
\footnotesize    
\caption{Population-level results based bootstrapped scores across 100 model runs. We report the average measures ($\pm std$) over 100 resampled bootstrap iterations. We calculate all measures over the positive class (unsustainable articles) besides Macro-F1 and AU-ROC. `AU-ROC' is the Area under the Receiver-Operator Curve. `Prec@X' is the Precision at X percentage of the most likely unsustainable articles by the model. The baseline model uses a single explanatory feature -- Num-of-Revisions-Normalized. $\uparrow$ Indicates that a higher value is better.}
{
\begin{tabular}{l@{\quad}l@{\quad}|c@{\quad}c@{\quad}c@{\quad}c@{\quad}c@{\quad}|c@{\quad}@{\quad}c@{}}
      {Type} & {Feature Set} & {Prec@2\% ($\uparrow$)} & {Prec@10\% ($\uparrow$)} & {Precision ($\uparrow$)} & {Recall ($\uparrow$)} & {F1 ($\uparrow$)}& {Macro F1 ($\uparrow$)} & {AU-ROC ($\uparrow$)}\\[2pt]
      \hline\rule{0pt}{12pt}
        \multirow{8}{*}{\rotatebox[origin=c]{360}{\footnotesize $FA$}} &
        Baseline                           & 0.61$\pm$0.09& 0.54$\pm$0.05& 0.31$\pm$0.03& 0.43$\pm$0.14& 0.35$\pm$0.03& 0.58$\pm$0.01& 0.67$\pm$0.02 \\[2pt]
    \cdashline{2-9} \rule{0pt}{12pt} &
    Network                                 & 0.54$\pm$0.08& 0.44$\pm$0.05& 0.33$\pm$0.02& 0.39$\pm$0.03& 0.36$\pm$0.02& 0.58$\pm$0.01& 0.66$\pm$0.02 \\[2pt] &
    Topics                                  & 0.75$\pm$0.08& 0.61$\pm$0.05& 0.46$\pm$0.04& 0.39$\pm$0.06& 0.42$\pm$0.03& 0.64$\pm$0.01& 0.72$\pm$0.02 \\[2pt] &
    Team Compos.                            & 0.74$\pm$0.08& 0.61$\pm$0.06& 0.41$\pm$0.02& 0.50$\pm$0.04& 0.45$\pm$0.02& 0.65$\pm$0.01& 0.74$\pm$0.01 \\[2pt] &
    Discussions                             & 0.40$\pm$0.09& 0.41$\pm$0.06& 0.39$\pm$0.02& 0.57$\pm$0.05& 0.46$\pm$0.02& 0.65$\pm$0.01& 0.75$\pm$0.01 \\[2pt] &
    Edit History                            & 0.79$\pm$0.09& 0.67$\pm$0.05& 0.44$\pm$0.03& 0.57$\pm$0.06& 0.49$\pm$0.02& 0.67$\pm$0.01& 0.80$\pm$0.01 \\[2pt] &
    User Experience                         & 0.78$\pm$0.08& 0.72$\pm$0.05& 0.54$\pm$0.03& 0.62$\pm$0.04& 0.58$\pm$0.02& 0.73$\pm$0.01& 0.85$\pm$0.01 \\[2pt] &
    All                                     & \textbf{0.84}$\pm$0.07& \textbf{0.79}$\pm$0.04& \textbf{0.62}$\pm$0.03& \textbf{0.68}$\pm$0.04& \textbf{0.65}$\pm$0.02& \textbf{0.78}$\pm$0.01& \textbf{0.89}$\pm$0.01 \\[2pt]
    \hline\rule{0pt}{12pt}
    
    \multirow{9}{*}{\rotatebox[origin=c]{360}{\footnotesize $GA$}} &
        Baseline                           & 0.43$\pm$0.04& 0.24$\pm$0.02& 0.38$\pm$0.06& 0.13$\pm$0.09& 0.18$\pm$0.02& 0.57$\pm$0.01& 0.70$\pm$0.01 \\[2pt]
    \cdashline{2-9} \rule{0pt}{12pt} &
    Network                                 & 0.25$\pm$0.03& 0.21$\pm$0.02& 0.19$\pm$0.01& 0.25$\pm$0.02& 0.22$\pm$0.01& 0.56$\pm$0.01& 0.68$\pm$0.01\\[2pt] &
    Topics                                  & 0.46$\pm$0.04& 0.33$\pm$0.02& 0.30$\pm$0.04& 0.25$\pm$0.07& 0.27$\pm$0.03& 0.60$\pm$0.01& 0.71$\pm$0.01\\[2pt] &
    Team Compos.                            & 0.44$\pm$0.03& 0.32$\pm$0.02& 0.26$\pm$0.01& 0.28$\pm$0.02& 0.27$\pm$0.01& 0.60$\pm$0.01& 0.72$\pm$0.01\\[2pt] &
    Discussions                             & 0.31$\pm$0.03& 0.27$\pm$0.02& 0.21$\pm$0.02& 0.42$\pm$0.11& 0.27$\pm$0.03& 0.58$\pm$0.01& 0.75$\pm$0.01\\[2pt] &
    Edit History                           & 0.53$\pm$0.03& 0.37$\pm$0.02& 0.33$\pm$0.02& 0.30$\pm$0.03& 0.31$\pm$0.02& 0.63$\pm$0.01& 0.78$\pm$0.01\\[2pt] &
    User Experience                         & 0.65$\pm$0.04& 0.53$\pm$0.02& 0.46$\pm$0.02& 0.45$\pm$0.02& 0.45$\pm$0.02& 0.70$\pm$0.01& 0.84$\pm$0.01\\[2pt] &
    All                                     & \textbf{0.74}$\pm$0.03& \textbf{0.58}$\pm$0.02& \textbf{0.50}$\pm$0.03& \textbf{0.47}$\pm$0.03& \textbf{0.48}$\pm$0.02& \textbf{0.72}$\pm$0.01& \textbf{0.87}$\pm$0.01 \\[2pt]
    \end{tabular}
}
\label{table:full_bootstrap_res}
\end{table*}

\begin{table*}
\centering
\footnotesize    
\caption{Five-Fold-CV modeling results. We report the average measures ($\pm std$) over five-stratified-fold-CVs. We calculate all measures over the minority class (unsustainable articles) besides Macro-F1 and AU-ROC. `AU-ROC' is the Area under the Receiver-Operator Curve. `Prec@X' is the Precision at X percentage of the most likely unsustainable articles by the model. The baseline model uses a single explanatory feature -- Num-of-Revisions-Normalized. $\uparrow$ Indicates that a higher value is better.}
{
\begin{tabular}{l@{\quad}l@{\quad}|c@{\quad}c@{\quad}c@{\quad}c@{\quad}c@{\quad}|c@{\quad}@{\quad}c@{}}
      {Type} & {Feature Set} & {Prec@2\% ($\uparrow$)} & {Prec@10\% ($\uparrow$)} & {Precision ($\uparrow$)} & {Recall ($\uparrow$)} & {F1 ($\uparrow$)}& {Macro F1 ($\uparrow$)} & {AU-ROC ($\uparrow$)}\\[2pt]
      \hline\rule{0pt}{12pt}
        \multirow{8}{*}{\rotatebox[origin=c]{360}{\footnotesize $FA$}} &
        Baseline                           & 0.66$\pm$0.13& 0.55$\pm$0.06& 0.31$\pm$0.02& 0.47$\pm$0.14& 0.36$\pm$0.03& 0.58$\pm$0.01& 0.67$\pm$0.01 \\[2pt]
    \cdashline{2-9} \rule{0pt}{12pt} &
    Network                                 & 0.55$\pm$0.02& 0.44$\pm$0.02& 0.32$\pm$0.03& 0.38$\pm$0.01& 0.35$\pm$0.02& 0.58$\pm$0.02& 0.66$\pm$0.02\\[2pt] &
    Topics                                  & 0.71$\pm$0.03& 0.61$\pm$0.03& 0.46$\pm$0.05& 0.39$\pm$0.04& 0.42$\pm$0.02& 0.64$\pm$0.01& 0.72$\pm$0.01\\[2pt] &
    Team Compos.                            & 0.70$\pm$0.10& 0.59$\pm$0.04& 0.42$\pm$0.03& 0.51$\pm$0.03& 0.46$\pm$0.02& 0.65$\pm$0.02& 0.74$\pm$0.01\\[2pt] &
    Discussions                             & 0.33$\pm$0.06& 0.35$\pm$0.04& 0.39$\pm$0.03& 0.52$\pm$0.06& 0.44$\pm$0.02& 0.64$\pm$0.01& 0.75$\pm$0.01\\[2pt] &
    Edit History                            & 0.77$\pm$0.06& 0.7$\pm$0.03& 0.44$\pm$0.03& 0.56$\pm$0.05& 0.49$\pm$0.02& 0.67$\pm$0.01& 0.80$\pm$0.01\\[2pt] &
    User Experience                         & 0.75$\pm$0.09& 0.71$\pm$0.04& 0.54$\pm$0.01& 0.63$\pm$0.05& 0.58$\pm$0.02& 0.72$\pm$0.01& 0.85$\pm$0.01\\[2pt] &
    All                                     & \textbf{0.80}$\pm$0.08& \textbf{0.77}$\pm$0.03& \textbf{0.62}$\pm$0.02& \textbf{0.67}$\pm$0.03& \textbf{0.65}$\pm$0.02& \textbf{0.77}$\pm$0.01& \textbf{0.90}$\pm$0.01 \\[2pt]
    \hline\rule{0pt}{12pt}
    
    \multirow{9}{*}{\rotatebox[origin=c]{360}{\footnotesize $GA$}} &
        Baseline                           & 0.42$\pm$0.02& 0.25$\pm$0.02& 0.39$\pm$0.03& 0.12$\pm$0.01& 0.18$\pm$0.01& 0.57$\pm$0.01& 0.70$\pm$0.01 \\[2pt]
    \cdashline{2-9} \rule{0pt}{12pt} &
    Network                                 & 0.23$\pm$0.04& 0.21$\pm$0.02& 0.19$\pm$0.01& 0.24$\pm$0.03& 0.21$\pm$0.02& 0.55$\pm$0.01& 0.68$\pm$0.01\\[2pt] &
    Topics                                  & 0.47$\pm$0.02& 0.33$\pm$0.01& 0.34$\pm$0.01& 0.20$\pm$0.01& 0.25$\pm$0.01& 0.60$\pm$0.01& 0.71$\pm$0.01\\[2pt] &
    Team Compos.                            & 0.45$\pm$0.02& 0.32$\pm$0.02& 0.26$\pm$0.01& 0.28$\pm$0.01& 0.27$\pm$0.01& 0.61$\pm$0.01& 0.71$\pm$0.01\\[2pt] &
    Discussions                             & 0.32$\pm$0.05& 0.27$\pm$0.02& 0.21$\pm$0.02& 0.41$\pm$0.13& 0.27$\pm$0.04& 0.58$\pm$0.01& 0.75$\pm$0.01\\[2pt] &
    Edit History                           & 0.53$\pm$0.05& 0.38$\pm$0.02& 0.33$\pm$0.02& 0.30$\pm$0.02& 0.31$\pm$0.02& 0.63$\pm$0.01& 0.78$\pm$0.02\\[2pt] &
    User Experience                         & 0.65$\pm$0.03& 0.53$\pm$0.02& 0.46$\pm$0.02& 0.45$\pm$0.03& 0.45$\pm$0.02& 0.70$\pm$0.01& 0.83$\pm$0.01\\[2pt] &
    All                                     & \textbf{0.75}$\pm$0.04& \textbf{0.57}$\pm$0.02& \textbf{0.48}$\pm$0.03& \textbf{0.47}$\pm$0.02& \textbf{0.48}$\pm$0.02& \textbf{0.72}$\pm$0.01& \textbf{0.87}$\pm$0.01 \\[2pt]
    \end{tabular}
}
\label{table:full_5_fold_cv_res}
\end{table*}

\clearpage

\begin{table}
\centering
\normalsize
\caption{The top ten False Positive (FP) articles. 
These articles have the highest unsustainability (demotion) potential according to our model (i.e., articles are sorted by the predicted likelihood of the model). 
`Pred.' represents the model's predicted likelihood that the article will become unsustainable.}
{
    \begin{tabular}{l@{\quad}l@{\quad}c@{\quad}|c@{}}
     \multicolumn{1}{c}{Article ID} &
     \multicolumn{1}{c}{Article Name} & 
     \multicolumn{1}{l|}{Pred.} & 
     \multicolumn{1}{c}{Reviews}\\[2pt]
      \hline\rule{0pt}{6pt}1797946     & 1928 Okeechobee hurricane         & 0.93 & 1\\[2pt]
        14533       & India                             & 0.92 & 3\\[2pt]
        25433       & Ronald Reagan                     & 0.92 & 2\\[2pt]
        186729      & Darjeeling                        & 0.91 & 2\\[2pt]
        58376       & École Polytechnique massacre      & 0.90 & 0\\[2pt]
        679074      & Quatermass and the Pit            & 0.90 & 1\\[2pt]
        2067        & Ann Arbor, Michigan               & 0.90 & \hspace{0.5em}2$^*$\\[2pt]
        27790       & Schizophrenia                     & 0.89 & 3\\[2pt]
        1267838     & Fightin' Texas Aggie Band         & 0.89 & 0\\[2pt]
        265965      & Chariot racing                    & 0.88 & 2\\[2pt]
    \hline
    \multicolumn{4}{l}{\footnotesize {$^*$Currently under a third review.}}\\
    \end{tabular}
    \label{table:top_10_false_positive}
}
\end{table}

\begin{table}
\centering
\normalsize
\caption{Articles that have been recently promoted to an $FA$ status (01/2019-09/2023), which have the highest potential to be unsustainable according to our model (i.e., articles are sorted by the predicted likelihood of the model). `Pred.' represents the model's predicted likelihood that the article will become unsustainable.}
{
    \begin{tabular}{l@{\quad}l@{\quad}l@{\quad}|c@{}}
     \multicolumn{1}{c}{Article ID} &
     \multicolumn{1}{c}{Article Name} & 
     \multicolumn{1}{l|}{$t_{prom.}$} & 
     \multicolumn{1}{c}{Pred.}\\[2pt]
      \hline\rule{0pt}{6pt}45051683    & Baby Driver                           & 11/2019   & 0.47  \\[2pt]
        67512396    & Yuzuru Hanyu Olympic seasons          & 10/2022   & 0.46  \\[2pt]
        28222625    & Sega                                  & 05/2020   & 0.45  \\[2pt]
        55728902    & Delicate (Taylor Swift song)          & 0.3/2021  & 0.43  \\[2pt]
        54611700    & Bluey (2018 TV series)                & 11/2020   & 0.42  \\[2pt]
        59729347    & 2020 Masters (snooker)                & 07/2020   & 0.28  \\[2pt]
        63109124    & 44th Chess Olympiad                   & 07/2023   & 0.27  \\[2pt]
        187509      & Pomona College                        & 12/2021   & 0.26  \\[2pt]
        71385873    & John Raymond$^{\#}$                   & 10/2022   & 0.25  \\[2pt]
        44627124    & Style (Taylor Swift song)             & 07/2021   & 0.24  \\[2pt]
    \hline
    \multicolumn{4}{l}{\footnotesize {$^{\#}$Full article name is `John Raymond science fiction magazines.'}}\\
    \end{tabular}
\label{table:demotion_potential_from_latests_years_promoted}
}
\end{table}

\hspace{0pt}
\begin{figure}
    \centering
    \includegraphics[scale=0.46]{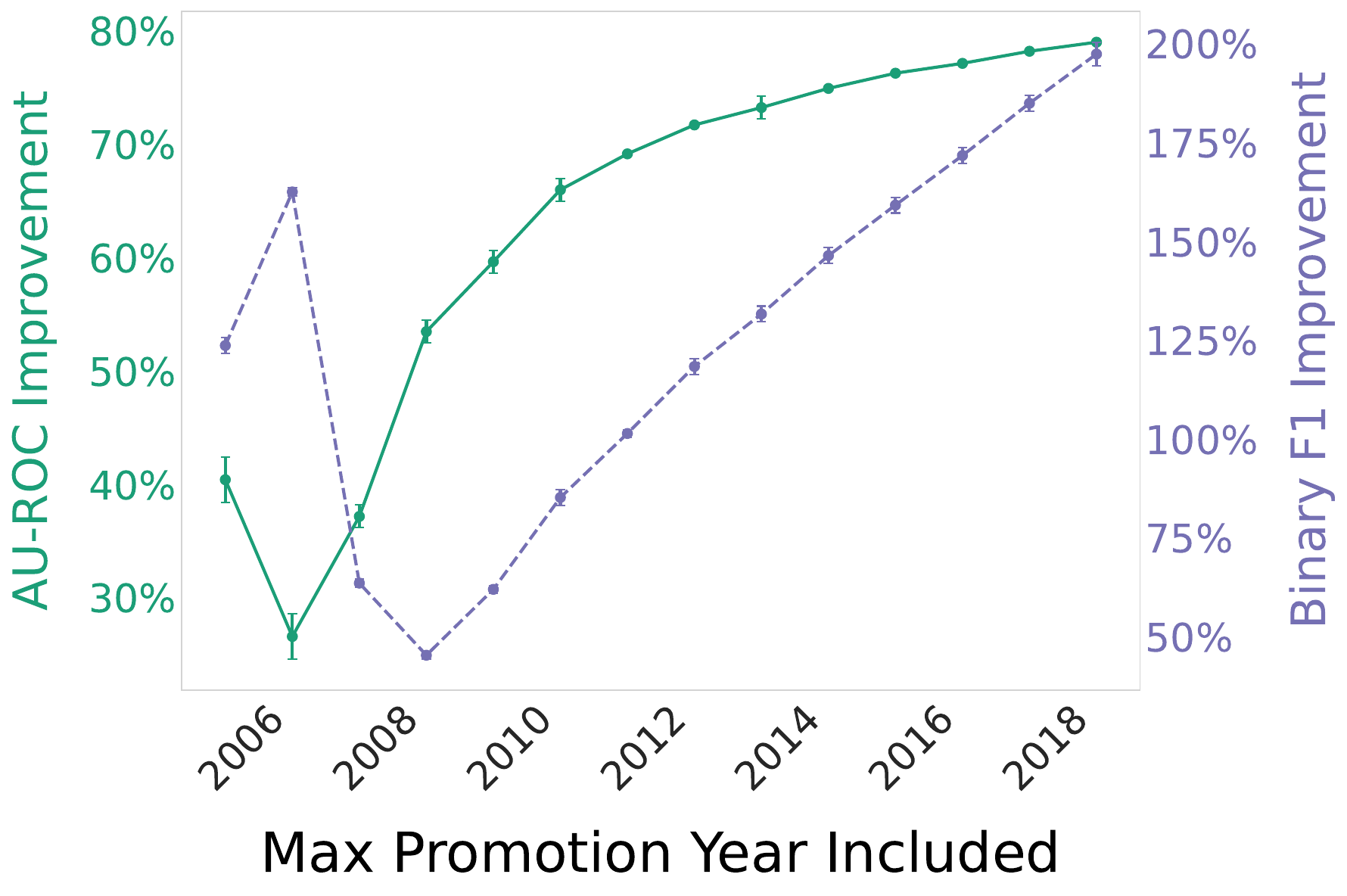}
    \caption{The model's performance improves as the corpus size gets larger and the model is exposed to more positive (unsustainable) articles. Performance improvement continues to increase and has not saturated. Improvement is measured as a percentage compared to a random model. As the corpus expands, the model shows better results in both measures (right and left y-axis). The dashed line corresponds to values on the right y-axis. While error bars are included, some are too small to be visually noticeable.}
    \label{fig:corpus_size_impacts_performance}
\end{figure}